\documentclass[longauth]{aa}

\usepackage{graphicx}
\usepackage{natbib}
\usepackage{scalerel}
\usepackage{multirow}   

\usepackage[table]{xcolor}

\usepackage{txfonts}
\usepackage[pdfencoding=auto,psdextra]{hyperref}
\hypersetup{
    colorlinks=true,
    linkcolor=blue,
    filecolor=magenta,      
    urlcolor=blue,
    citecolor=blue
}
\makeatletter
\renewcommand*\aa@pageof{, page \thepage{} of \pageref*{LastPage}}
\makeatother

\usepackage[utf8]{inputenc}

\usepackage{euclid}
\usepackage{tabularx}

\renewcommand{\linenumbers}{}

\begin{document}
%
%

\title{\Euclid preparation}
\subtitle{Impact of redshift distribution uncertainties on the joint analysis of photometric galaxy clustering and weak gravitational lensing}    
   
\newcommand{\orcid}[1]{} 

%
%
%
%

\renewcommand{\orcid}[1]{} 
\author{Euclid Collaboration: K.~A.~Bertmann\orcid{0009-0004-6700-2470}\thanks{\email{klara.bertmann@astro.ruhr-uni-bochum.de}}\inst{\ref{aff1}}
\and A.~Porredon\orcid{0000-0002-2762-2024}\inst{\ref{aff2},\ref{aff1}}
\and V.~Duret\orcid{0009-0009-0383-4960}\inst{\ref{aff3}}
\and J.~Fonseca\orcid{0000-0003-0549-1614}\inst{\ref{aff4},\ref{aff5},\ref{aff6}}
\and H.~Hildebrandt\orcid{0000-0002-9814-3338}\inst{\ref{aff1}}
\and I.~Tutusaus\orcid{0000-0002-3199-0399}\inst{\ref{aff7},\ref{aff8},\ref{aff9}}
\and S.~Camera\orcid{0000-0003-3399-3574}\inst{\ref{aff10},\ref{aff11},\ref{aff12}}
\and S.~Escoffier\orcid{0000-0002-2847-7498}\inst{\ref{aff3}}
\and N.~Aghanim\orcid{0000-0002-6688-8992}\inst{\ref{aff13}}
\and B.~Altieri\orcid{0000-0003-3936-0284}\inst{\ref{aff14}}
\and A.~Amara\inst{\ref{aff15}}
\and S.~Andreon\orcid{0000-0002-2041-8784}\inst{\ref{aff16}}
\and N.~Auricchio\orcid{0000-0003-4444-8651}\inst{\ref{aff17}}
\and C.~Baccigalupi\orcid{0000-0002-8211-1630}\inst{\ref{aff18},\ref{aff19},\ref{aff20},\ref{aff21}}
\and M.~Baldi\orcid{0000-0003-4145-1943}\inst{\ref{aff22},\ref{aff17},\ref{aff23}}
\and S.~Bardelli\orcid{0000-0002-8900-0298}\inst{\ref{aff17}}
\and P.~Battaglia\orcid{0000-0002-7337-5909}\inst{\ref{aff17}}
\and A.~Biviano\orcid{0000-0002-0857-0732}\inst{\ref{aff19},\ref{aff18}}
\and E.~Branchini\orcid{0000-0002-0808-6908}\inst{\ref{aff24},\ref{aff25},\ref{aff16}}
\and M.~Brescia\orcid{0000-0001-9506-5680}\inst{\ref{aff26},\ref{aff27}}
\and G.~Ca\~nas-Herrera\orcid{0000-0003-2796-2149}\inst{\ref{aff28},\ref{aff29}}
\and V.~Capobianco\orcid{0000-0002-3309-7692}\inst{\ref{aff12}}
\and C.~Carbone\orcid{0000-0003-0125-3563}\inst{\ref{aff30}}
\and V.~F.~Cardone\inst{\ref{aff31},\ref{aff32}}
\and J.~Carretero\orcid{0000-0002-3130-0204}\inst{\ref{aff2},\ref{aff33}}
\and S.~Casas\orcid{0000-0002-4751-5138}\inst{\ref{aff34},\ref{aff35}}
\and F.~J.~Castander\orcid{0000-0001-7316-4573}\inst{\ref{aff7},\ref{aff8}}
\and M.~Castellano\orcid{0000-0001-9875-8263}\inst{\ref{aff31}}
\and G.~Castignani\orcid{0000-0001-6831-0687}\inst{\ref{aff17}}
\and S.~Cavuoti\orcid{0000-0002-3787-4196}\inst{\ref{aff27},\ref{aff36}}
\and K.~C.~Chambers\orcid{0000-0001-6965-7789}\inst{\ref{aff37}}
\and A.~Cimatti\inst{\ref{aff38}}
\and C.~Colodro-Conde\inst{\ref{aff39}}
\and G.~Congedo\orcid{0000-0003-2508-0046}\inst{\ref{aff28}}
\and L.~Conversi\orcid{0000-0002-6710-8476}\inst{\ref{aff40},\ref{aff14}}
\and Y.~Copin\orcid{0000-0002-5317-7518}\inst{\ref{aff41}}
\and F.~Courbin\orcid{0000-0003-0758-6510}\inst{\ref{aff42},\ref{aff43},\ref{aff44}}
\and H.~M.~Courtois\orcid{0000-0003-0509-1776}\inst{\ref{aff45}}
\and M.~Cropper\orcid{0000-0003-4571-9468}\inst{\ref{aff46}}
\and A.~Da~Silva\orcid{0000-0002-6385-1609}\inst{\ref{aff47},\ref{aff48}}
\and H.~Degaudenzi\orcid{0000-0002-5887-6799}\inst{\ref{aff49}}
\and G.~De~Lucia\orcid{0000-0002-6220-9104}\inst{\ref{aff19}}
\and H.~Dole\orcid{0000-0002-9767-3839}\inst{\ref{aff13}}
\and M.~Douspis\orcid{0000-0003-4203-3954}\inst{\ref{aff13}}
\and F.~Dubath\orcid{0000-0002-6533-2810}\inst{\ref{aff49}}
\and X.~Dupac\inst{\ref{aff14}}
\and S.~Dusini\orcid{0000-0002-1128-0664}\inst{\ref{aff50}}
\and M.~Farina\orcid{0000-0002-3089-7846}\inst{\ref{aff51}}
\and R.~Farinelli\inst{\ref{aff17}}
\and S.~Farrens\orcid{0000-0002-9594-9387}\inst{\ref{aff52}}
\and S.~Ferriol\inst{\ref{aff41}}
\and F.~Finelli\orcid{0000-0002-6694-3269}\inst{\ref{aff17},\ref{aff53}}
\and P.~Fosalba\orcid{0000-0002-1510-5214}\inst{\ref{aff8},\ref{aff7}}
\and S.~Fotopoulou\orcid{0000-0002-9686-254X}\inst{\ref{aff54}}
\and N.~Fourmanoit\orcid{0009-0005-6816-6925}\inst{\ref{aff3}}
\and M.~Frailis\orcid{0000-0002-7400-2135}\inst{\ref{aff19}}
\and E.~Franceschi\orcid{0000-0002-0585-6591}\inst{\ref{aff17}}
\and M.~Fumana\orcid{0000-0001-6787-5950}\inst{\ref{aff30}}
\and S.~Galeotta\orcid{0000-0002-3748-5115}\inst{\ref{aff19}}
\and K.~George\orcid{0000-0002-1734-8455}\inst{\ref{aff55}}
\and W.~Gillard\orcid{0000-0003-4744-9748}\inst{\ref{aff3}}
\and B.~Gillis\orcid{0000-0002-4478-1270}\inst{\ref{aff28}}
\and C.~Giocoli\orcid{0000-0002-9590-7961}\inst{\ref{aff17},\ref{aff23}}
\and J.~Gracia-Carpio\orcid{0000-0003-4689-3134}\inst{\ref{aff56}}
\and A.~Grazian\orcid{0000-0002-5688-0663}\inst{\ref{aff57}}
\and F.~Grupp\inst{\ref{aff56},\ref{aff58}}
\and S.~V.~H.~Haugan\orcid{0000-0001-9648-7260}\inst{\ref{aff59}}
\and H.~Hoekstra\orcid{0000-0002-0641-3231}\inst{\ref{aff29}}
\and W.~Holmes\inst{\ref{aff60}}
\and F.~Hormuth\inst{\ref{aff61}}
\and A.~Hornstrup\orcid{0000-0002-3363-0936}\inst{\ref{aff62},\ref{aff63}}
\and K.~Jahnke\orcid{0000-0003-3804-2137}\inst{\ref{aff64}}
\and M.~Jhabvala\inst{\ref{aff65}}
\and B.~Joachimi\orcid{0000-0001-7494-1303}\inst{\ref{aff66}}
\and S.~Kermiche\orcid{0000-0002-0302-5735}\inst{\ref{aff3}}
\and A.~Kiessling\orcid{0000-0002-2590-1273}\inst{\ref{aff60}}
\and M.~Kilbinger\orcid{0000-0001-9513-7138}\inst{\ref{aff52}}
\and B.~Kubik\orcid{0009-0006-5823-4880}\inst{\ref{aff41}}
\and M.~Kunz\orcid{0000-0002-3052-7394}\inst{\ref{aff67}}
\and H.~Kurki-Suonio\orcid{0000-0002-4618-3063}\inst{\ref{aff68},\ref{aff69}}
\and A.~M.~C.~Le~Brun\orcid{0000-0002-0936-4594}\inst{\ref{aff70}}
\and S.~Ligori\orcid{0000-0003-4172-4606}\inst{\ref{aff12}}
\and P.~B.~Lilje\orcid{0000-0003-4324-7794}\inst{\ref{aff59}}
\and V.~Lindholm\orcid{0000-0003-2317-5471}\inst{\ref{aff68},\ref{aff69}}
\and I.~Lloro\orcid{0000-0001-5966-1434}\inst{\ref{aff71}}
\and G.~Mainetti\orcid{0000-0003-2384-2377}\inst{\ref{aff72}}
\and D.~Maino\inst{\ref{aff73},\ref{aff30},\ref{aff74}}
\and E.~Maiorano\orcid{0000-0003-2593-4355}\inst{\ref{aff17}}
\and O.~Mansutti\orcid{0000-0001-5758-4658}\inst{\ref{aff19}}
\and S.~Marcin\inst{\ref{aff75}}
\and O.~Marggraf\orcid{0000-0001-7242-3852}\inst{\ref{aff76}}
\and M.~Martinelli\orcid{0000-0002-6943-7732}\inst{\ref{aff31},\ref{aff32}}
\and N.~Martinet\orcid{0000-0003-2786-7790}\inst{\ref{aff77}}
\and F.~Marulli\orcid{0000-0002-8850-0303}\inst{\ref{aff78},\ref{aff17},\ref{aff23}}
\and R.~J.~Massey\orcid{0000-0002-6085-3780}\inst{\ref{aff79}}
\and E.~Medinaceli\orcid{0000-0002-4040-7783}\inst{\ref{aff17}}
\and S.~Mei\orcid{0000-0002-2849-559X}\inst{\ref{aff80},\ref{aff81}}
\and Y.~Mellier\thanks{Deceased}\inst{\ref{aff82},\ref{aff83}}
\and M.~Meneghetti\orcid{0000-0003-1225-7084}\inst{\ref{aff17},\ref{aff23}}
\and E.~Merlin\orcid{0000-0001-6870-8900}\inst{\ref{aff31}}
\and G.~Meylan\inst{\ref{aff84}}
\and A.~Mora\orcid{0000-0002-1922-8529}\inst{\ref{aff85}}
\and M.~Moresco\orcid{0000-0002-7616-7136}\inst{\ref{aff78},\ref{aff17}}
\and L.~Moscardini\orcid{0000-0002-3473-6716}\inst{\ref{aff78},\ref{aff17},\ref{aff23}}
\and R.~Nakajima\orcid{0009-0009-1213-7040}\inst{\ref{aff76}}
\and C.~Neissner\orcid{0000-0001-8524-4968}\inst{\ref{aff86},\ref{aff33}}
\and S.-M.~Niemi\orcid{0009-0005-0247-0086}\inst{\ref{aff87}}
\and J.~W.~Nightingale\orcid{0000-0002-8987-7401}\inst{\ref{aff88}}
\and C.~Padilla\orcid{0000-0001-7951-0166}\inst{\ref{aff86}}
\and S.~Paltani\orcid{0000-0002-8108-9179}\inst{\ref{aff49}}
\and F.~Pasian\orcid{0000-0002-4869-3227}\inst{\ref{aff19}}
\and K.~Pedersen\inst{\ref{aff89}}
\and W.~J.~Percival\orcid{0000-0002-0644-5727}\inst{\ref{aff90},\ref{aff91},\ref{aff92}}
\and V.~Pettorino\orcid{0000-0002-4203-9320}\inst{\ref{aff87}}
\and G.~Polenta\orcid{0000-0003-4067-9196}\inst{\ref{aff93}}
\and M.~Poncet\inst{\ref{aff94}}
\and L.~A.~Popa\inst{\ref{aff95}}
\and L.~Pozzetti\orcid{0000-0001-7085-0412}\inst{\ref{aff17}}
\and F.~Raison\orcid{0000-0002-7819-6918}\inst{\ref{aff56}}
\and A.~Renzi\orcid{0000-0001-9856-1970}\inst{\ref{aff96},\ref{aff50}}
\and J.~Rhodes\orcid{0000-0002-4485-8549}\inst{\ref{aff60}}
\and G.~Riccio\inst{\ref{aff27}}
\and E.~Romelli\orcid{0000-0003-3069-9222}\inst{\ref{aff19}}
\and M.~Roncarelli\orcid{0000-0001-9587-7822}\inst{\ref{aff17}}
\and C.~Rosset\orcid{0000-0003-0286-2192}\inst{\ref{aff80}}
\and R.~Saglia\orcid{0000-0003-0378-7032}\inst{\ref{aff58},\ref{aff56}}
\and Z.~Sakr\orcid{0000-0002-4823-3757}\inst{\ref{aff97},\ref{aff9},\ref{aff98}}
\and D.~Sapone\orcid{0000-0001-7089-4503}\inst{\ref{aff99}}
\and B.~Sartoris\orcid{0000-0003-1337-5269}\inst{\ref{aff58},\ref{aff19}}
\and P.~Schneider\orcid{0000-0001-8561-2679}\inst{\ref{aff76}}
\and T.~Schrabback\orcid{0000-0002-6987-7834}\inst{\ref{aff100}}
\and A.~Secroun\orcid{0000-0003-0505-3710}\inst{\ref{aff3}}
\and E.~Sefusatti\orcid{0000-0003-0473-1567}\inst{\ref{aff19},\ref{aff18},\ref{aff20}}
\and G.~Seidel\orcid{0000-0003-2907-353X}\inst{\ref{aff64}}
\and E.~Sihvola\orcid{0000-0003-1804-7715}\inst{\ref{aff101}}
\and P.~Simon\inst{\ref{aff76}}
\and C.~Sirignano\orcid{0000-0002-0995-7146}\inst{\ref{aff96},\ref{aff50}}
\and G.~Sirri\orcid{0000-0003-2626-2853}\inst{\ref{aff23}}
\and A.~Spurio~Mancini\orcid{0000-0001-5698-0990}\inst{\ref{aff102}}
\and L.~Stanco\orcid{0000-0002-9706-5104}\inst{\ref{aff50}}
\and P.~Tallada-Cresp\'{i}\orcid{0000-0002-1336-8328}\inst{\ref{aff2},\ref{aff33}}
\and A.~N.~Taylor\inst{\ref{aff28}}
\and I.~Tereno\orcid{0000-0002-4537-6218}\inst{\ref{aff47},\ref{aff103}}
\and N.~Tessore\orcid{0000-0002-9696-7931}\inst{\ref{aff46}}
\and S.~Toft\orcid{0000-0003-3631-7176}\inst{\ref{aff104},\ref{aff105}}
\and R.~Toledo-Moreo\orcid{0000-0002-2997-4859}\inst{\ref{aff106}}
\and F.~Torradeflot\orcid{0000-0003-1160-1517}\inst{\ref{aff33},\ref{aff2}}
\and J.~Valiviita\orcid{0000-0001-6225-3693}\inst{\ref{aff68},\ref{aff69}}
\and T.~Vassallo\orcid{0000-0001-6512-6358}\inst{\ref{aff19}}
\and G.~Verdoes~Kleijn\orcid{0000-0001-5803-2580}\inst{\ref{aff107}}
\and Y.~Wang\orcid{0000-0002-4749-2984}\inst{\ref{aff108}}
\and J.~Weller\orcid{0000-0002-8282-2010}\inst{\ref{aff58},\ref{aff56}}
\and G.~Zamorani\orcid{0000-0002-2318-301X}\inst{\ref{aff17}}
\and F.~M.~Zerbi\inst{\ref{aff16}}
\and E.~Zucca\orcid{0000-0002-5845-8132}\inst{\ref{aff17}}
\and M.~Ballardini\orcid{0000-0003-4481-3559}\inst{\ref{aff109},\ref{aff110},\ref{aff17}}
\and A.~Boucaud\orcid{0000-0001-7387-2633}\inst{\ref{aff80}}
\and E.~Bozzo\orcid{0000-0002-8201-1525}\inst{\ref{aff49}}
\and C.~Burigana\orcid{0000-0002-3005-5796}\inst{\ref{aff111},\ref{aff53}}
\and R.~Cabanac\orcid{0000-0001-6679-2600}\inst{\ref{aff9}}
\and M.~Calabrese\orcid{0000-0002-2637-2422}\inst{\ref{aff112},\ref{aff30}}
\and A.~Cappi\inst{\ref{aff113},\ref{aff17}}
\and T.~Castro\orcid{0000-0002-6292-3228}\inst{\ref{aff19},\ref{aff20},\ref{aff18},\ref{aff114}}
\and J.~A.~Escartin~Vigo\inst{\ref{aff56}}
\and G.~Fabbian\orcid{0000-0002-3255-4695}\inst{\ref{aff13}}
\and J.~Macias-Perez\orcid{0000-0002-5385-2763}\inst{\ref{aff115}}
\and R.~Maoli\orcid{0000-0002-6065-3025}\inst{\ref{aff116},\ref{aff31}}
\and J.~Mart\'{i}n-Fleitas\orcid{0000-0002-8594-569X}\inst{\ref{aff117}}
\and N.~Mauri\orcid{0000-0001-8196-1548}\inst{\ref{aff38},\ref{aff23}}
\and R.~B.~Metcalf\orcid{0000-0003-3167-2574}\inst{\ref{aff78},\ref{aff17}}
\and P.~Monaco\orcid{0000-0003-2083-7564}\inst{\ref{aff118},\ref{aff19},\ref{aff20},\ref{aff18}}
\and A.~A.~Nucita\inst{\ref{aff119},\ref{aff120},\ref{aff121}}
\and A.~Pezzotta\orcid{0000-0003-0726-2268}\inst{\ref{aff16}}
\and M.~P\"ontinen\orcid{0000-0001-5442-2530}\inst{\ref{aff68}}
\and I.~Risso\orcid{0000-0003-2525-7761}\inst{\ref{aff16},\ref{aff25}}
\and V.~Scottez\orcid{0009-0008-3864-940X}\inst{\ref{aff82},\ref{aff122}}
\and M.~Sereno\orcid{0000-0003-0302-0325}\inst{\ref{aff17},\ref{aff23}}
\and M.~Tenti\orcid{0000-0002-4254-5901}\inst{\ref{aff23}}
\and M.~Tucci\inst{\ref{aff49}}
\and M.~Viel\orcid{0000-0002-2642-5707}\inst{\ref{aff18},\ref{aff19},\ref{aff21},\ref{aff20},\ref{aff114}}
\and M.~Wiesmann\orcid{0009-0000-8199-5860}\inst{\ref{aff59}}
\and Y.~Akrami\orcid{0000-0002-2407-7956}\inst{\ref{aff123},\ref{aff124}}
\and I.~T.~Andika\orcid{0000-0001-6102-9526}\inst{\ref{aff55}}
\and G.~Angora\orcid{0000-0002-0316-6562}\inst{\ref{aff27},\ref{aff109}}
\and S.~Anselmi\orcid{0000-0002-3579-9583}\inst{\ref{aff50},\ref{aff96},\ref{aff125}}
\and M.~Archidiacono\orcid{0000-0003-4952-9012}\inst{\ref{aff73},\ref{aff74}}
\and F.~Atrio-Barandela\orcid{0000-0002-2130-2513}\inst{\ref{aff126}}
\and L.~Bazzanini\orcid{0000-0003-0727-0137}\inst{\ref{aff109},\ref{aff17}}
\and D.~Bertacca\orcid{0000-0002-2490-7139}\inst{\ref{aff96},\ref{aff57},\ref{aff50}}
\and M.~Bethermin\orcid{0000-0002-3915-2015}\inst{\ref{aff127}}
\and F.~Beutler\orcid{0000-0003-0467-5438}\inst{\ref{aff28}}
\and A.~Blanchard\orcid{0000-0001-8555-9003}\inst{\ref{aff9}}
\and L.~Blot\orcid{0000-0002-9622-7167}\inst{\ref{aff128},\ref{aff70}}
\and M.~Bonici\orcid{0000-0002-8430-126X}\inst{\ref{aff90},\ref{aff30}}
\and S.~Borgani\orcid{0000-0001-6151-6439}\inst{\ref{aff118},\ref{aff18},\ref{aff19},\ref{aff20},\ref{aff114}}
\and M.~L.~Brown\orcid{0000-0002-0370-8077}\inst{\ref{aff129}}
\and S.~Bruton\orcid{0000-0002-6503-5218}\inst{\ref{aff130}}
\and A.~Calabro\orcid{0000-0003-2536-1614}\inst{\ref{aff31}}
\and B.~Camacho~Quevedo\orcid{0000-0002-8789-4232}\inst{\ref{aff18},\ref{aff21},\ref{aff19}}
\and F.~Caro\inst{\ref{aff31}}
\and C.~S.~Carvalho\inst{\ref{aff103}}
\and F.~Cogato\orcid{0000-0003-4632-6113}\inst{\ref{aff78},\ref{aff17}}
\and S.~Conseil\orcid{0000-0002-3657-4191}\inst{\ref{aff41}}
\and A.~R.~Cooray\orcid{0000-0002-3892-0190}\inst{\ref{aff131}}
\and S.~Davini\orcid{0000-0003-3269-1718}\inst{\ref{aff25}}
\and G.~Desprez\orcid{0000-0001-8325-1742}\inst{\ref{aff107}}
\and A.~D\'iaz-S\'anchez\orcid{0000-0003-0748-4768}\inst{\ref{aff132}}
\and S.~Di~Domizio\orcid{0000-0003-2863-5895}\inst{\ref{aff24},\ref{aff25}}
\and J.~M.~Diego\orcid{0000-0001-9065-3926}\inst{\ref{aff133}}
\and M.~Y.~Elkhashab\orcid{0000-0001-9306-2603}\inst{\ref{aff19},\ref{aff20},\ref{aff118},\ref{aff18}}
\and A.~Enia\orcid{0000-0002-0200-2857}\inst{\ref{aff17}}
\and Y.~Fang\orcid{0000-0002-0334-6950}\inst{\ref{aff58}}
\and A.~G.~Ferrari\orcid{0009-0005-5266-4110}\inst{\ref{aff23}}
\and A.~Finoguenov\orcid{0000-0002-4606-5403}\inst{\ref{aff68}}
\and F.~Fontanot\orcid{0000-0003-4744-0188}\inst{\ref{aff19},\ref{aff18}}
\and A.~Franco\orcid{0000-0002-4761-366X}\inst{\ref{aff120},\ref{aff119},\ref{aff121}}
\and K.~Ganga\orcid{0000-0001-8159-8208}\inst{\ref{aff80}}
\and T.~Gasparetto\orcid{0000-0002-7913-4866}\inst{\ref{aff31}}
\and E.~Gaztanaga\orcid{0000-0001-9632-0815}\inst{\ref{aff7},\ref{aff8},\ref{aff134}}
\and F.~Giacomini\orcid{0000-0002-3129-2814}\inst{\ref{aff23}}
\and F.~Gianotti\orcid{0000-0003-4666-119X}\inst{\ref{aff17}}
\and G.~Gozaliasl\orcid{0000-0002-0236-919X}\inst{\ref{aff135},\ref{aff68}}
\and A.~Gruppuso\orcid{0000-0001-9272-5292}\inst{\ref{aff17},\ref{aff23}}
\and M.~Guidi\orcid{0000-0001-9408-1101}\inst{\ref{aff22},\ref{aff17}}
\and C.~M.~Gutierrez\orcid{0000-0001-7854-783X}\inst{\ref{aff39},\ref{aff136}}
\and A.~Hall\orcid{0000-0002-3139-8651}\inst{\ref{aff28}}
\and C.~Hern\'andez-Monteagudo\orcid{0000-0001-5471-9166}\inst{\ref{aff136},\ref{aff39}}
\and J.~Hjorth\orcid{0000-0002-4571-2306}\inst{\ref{aff89}}
\and S.~Joudaki\orcid{0000-0001-8820-673X}\inst{\ref{aff2},\ref{aff134}}
\and J.~J.~E.~Kajava\orcid{0000-0002-3010-8333}\inst{\ref{aff137},\ref{aff138},\ref{aff139}}
\and Y.~Kang\orcid{0009-0000-8588-7250}\inst{\ref{aff49}}
\and V.~Kansal\orcid{0000-0002-4008-6078}\inst{\ref{aff140},\ref{aff141}}
\and D.~Karagiannis\orcid{0000-0002-4927-0816}\inst{\ref{aff109},\ref{aff6}}
\and K.~Kiiveri\inst{\ref{aff101}}
\and J.~Kim\orcid{0000-0003-2776-2761}\inst{\ref{aff142}}
\and C.~C.~Kirkpatrick\inst{\ref{aff101}}
\and S.~Kruk\orcid{0000-0001-8010-8879}\inst{\ref{aff14}}
\and M.~Lattanzi\orcid{0000-0003-1059-2532}\inst{\ref{aff110}}
\and L.~Legrand\orcid{0000-0003-0610-5252}\inst{\ref{aff143},\ref{aff144}}
\and M.~Lembo\orcid{0000-0002-5271-5070}\inst{\ref{aff83}}
\and F.~Lepori\orcid{0009-0000-5061-7138}\inst{\ref{aff145}}
\and G.~Leroy\orcid{0009-0004-2523-4425}\inst{\ref{aff146},\ref{aff79}}
\and G.~F.~Lesci\orcid{0000-0002-4607-2830}\inst{\ref{aff78},\ref{aff17}}
\and J.~Lesgourgues\orcid{0000-0001-7627-353X}\inst{\ref{aff34}}
\and T.~I.~Liaudat\orcid{0000-0002-9104-314X}\inst{\ref{aff147}}
\and S.~J.~Liu\orcid{0000-0001-7680-2139}\inst{\ref{aff51}}
\and M.~Magliocchetti\orcid{0000-0001-9158-4838}\inst{\ref{aff51}}
\and A.~Manj\'on-Garc\'ia\orcid{0000-0002-7413-8825}\inst{\ref{aff132}}
\and F.~Mannucci\orcid{0000-0002-4803-2381}\inst{\ref{aff148}}
\and C.~J.~A.~P.~Martins\orcid{0000-0002-4886-9261}\inst{\ref{aff149},\ref{aff4}}
\and M.~Migliaccio\inst{\ref{aff150},\ref{aff151}}
\and M.~Miluzio\inst{\ref{aff14},\ref{aff152}}
\and A.~Montoro\orcid{0000-0003-4730-8590}\inst{\ref{aff7},\ref{aff8}}
\and C.~Moretti\orcid{0000-0003-3314-8936}\inst{\ref{aff19},\ref{aff18},\ref{aff20}}
\and G.~Morgante\inst{\ref{aff17}}
\and S.~Nadathur\orcid{0000-0001-9070-3102}\inst{\ref{aff134}}
\and A.~Navarro-Alsina\orcid{0000-0002-3173-2592}\inst{\ref{aff76}}
\and S.~Nesseris\orcid{0000-0002-0567-0324}\inst{\ref{aff123}}
\and L.~Pagano\orcid{0000-0003-1820-5998}\inst{\ref{aff109},\ref{aff110}}
\and D.~Paoletti\orcid{0000-0003-4761-6147}\inst{\ref{aff17},\ref{aff53}}
\and F.~Passalacqua\orcid{0000-0002-8606-4093}\inst{\ref{aff96},\ref{aff50}}
\and K.~Paterson\orcid{0000-0001-8340-3486}\inst{\ref{aff64}}
\and L.~Patrizii\inst{\ref{aff23}}
\and A.~Pisani\orcid{0000-0002-6146-4437}\inst{\ref{aff3}}
\and D.~Potter\orcid{0000-0002-0757-5195}\inst{\ref{aff145}}
\and G.~W.~Pratt\inst{\ref{aff52}}
\and S.~Quai\orcid{0000-0002-0449-8163}\inst{\ref{aff78},\ref{aff17}}
\and M.~Radovich\orcid{0000-0002-3585-866X}\inst{\ref{aff57}}
\and K.~Rojas\orcid{0000-0003-1391-6854}\inst{\ref{aff75}}
\and W.~Roster\orcid{0000-0002-9149-6528}\inst{\ref{aff56}}
\and S.~Sacquegna\orcid{0000-0002-8433-6630}\inst{\ref{aff153}}
\and D.~B.~Sanders\orcid{0000-0002-1233-9998}\inst{\ref{aff37}}
\and E.~Sarpa\orcid{0000-0002-1256-655X}\inst{\ref{aff21},\ref{aff114},\ref{aff19}}
\and A.~Schneider\orcid{0000-0001-7055-8104}\inst{\ref{aff145}}
\and M.~Schultheis\inst{\ref{aff113}}
\and D.~Sciotti\orcid{0009-0008-4519-2620}\inst{\ref{aff31},\ref{aff32}}
\and E.~Sellentin\inst{\ref{aff154},\ref{aff29}}
\and L.~C.~Smith\orcid{0000-0002-3259-2771}\inst{\ref{aff155}}
\and K.~Tanidis\orcid{0000-0001-9843-5130}\inst{\ref{aff156}}
\and C.~Tao\orcid{0000-0001-7961-8177}\inst{\ref{aff3}}
\and F.~Tarsitano\orcid{0000-0002-5919-0238}\inst{\ref{aff157},\ref{aff49}}
\and G.~Testera\inst{\ref{aff25}}
\and R.~Teyssier\orcid{0000-0001-7689-0933}\inst{\ref{aff158}}
\and S.~Tosi\orcid{0000-0002-7275-9193}\inst{\ref{aff24},\ref{aff16},\ref{aff25}}
\and A.~Troja\orcid{0000-0003-0239-4595}\inst{\ref{aff96},\ref{aff50}}
\and A.~Venhola\orcid{0000-0001-6071-4564}\inst{\ref{aff159}}
\and D.~Vergani\orcid{0000-0003-0898-2216}\inst{\ref{aff17}}
\and F.~Vernizzi\orcid{0000-0003-3426-2802}\inst{\ref{aff160}}
\and G.~Verza\orcid{0000-0002-1886-8348}\inst{\ref{aff161},\ref{aff162}}
\and S.~Vinciguerra\orcid{0009-0005-4018-3184}\inst{\ref{aff77}}
\and N.~A.~Walton\orcid{0000-0003-3983-8778}\inst{\ref{aff155}}
\and A.~H.~Wright\orcid{0000-0001-7363-7932}\inst{\ref{aff1}}}
										   
\institute{Ruhr University Bochum, Faculty of Physics and Astronomy, Astronomical Institute (AIRUB), German Centre for Cosmological Lensing (GCCL), 44780 Bochum, Germany\label{aff1}
\and
Centro de Investigaciones Energ\'eticas, Medioambientales y Tecnol\'ogicas (CIEMAT), Avenida Complutense 40, 28040 Madrid, Spain\label{aff2}
\and
Aix-Marseille Universit\'e, CNRS/IN2P3, CPPM, Marseille, France\label{aff3}
\and
Instituto de Astrof\'isica e Ci\^encias do Espa\c{c}o, Universidade do Porto, CAUP, Rua das Estrelas, PT4150-762 Porto, Portugal\label{aff4}
\and
Faculdade de Ci\^encias da Universidade do Porto, Rua do Campo de Alegre, 4150-007 Porto, Portugal\label{aff5}
\and
Department of Physics and Astronomy, University of the Western Cape, Bellville, Cape Town, 7535, South Africa\label{aff6}
\and
Institute of Space Sciences (ICE, CSIC), Campus UAB, Carrer de Can Magrans, s/n, 08193 Barcelona, Spain\label{aff7}
\and
Institut d'Estudis Espacials de Catalunya (IEEC),  Edifici RDIT, Campus UPC, 08860 Castelldefels, Barcelona, Spain\label{aff8}
\and
Institut de Recherche en Astrophysique et Plan\'etologie (IRAP), Universit\'e de Toulouse, CNRS, UPS, CNES, 14 Av. Edouard Belin, 31400 Toulouse, France\label{aff9}
\and
Dipartimento di Fisica, Universit\`a degli Studi di Torino, Via P. Giuria 1, 10125 Torino, Italy\label{aff10}
\and
INFN-Sezione di Torino, Via P. Giuria 1, 10125 Torino, Italy\label{aff11}
\and
INAF-Osservatorio Astrofisico di Torino, Via Osservatorio 20, 10025 Pino Torinese (TO), Italy\label{aff12}
\and
Universit\'e Paris-Saclay, CNRS, Institut d'astrophysique spatiale, 91405, Orsay, France\label{aff13}
\and
ESAC/ESA, Camino Bajo del Castillo, s/n., Urb. Villafranca del Castillo, 28692 Villanueva de la Ca\~nada, Madrid, Spain\label{aff14}
\and
School of Mathematics and Physics, University of Surrey, Guildford, Surrey, GU2 7XH, UK\label{aff15}
\and
INAF-Osservatorio Astronomico di Brera, Via Brera 28, 20122 Milano, Italy\label{aff16}
\and
INAF-Osservatorio di Astrofisica e Scienza dello Spazio di Bologna, Via Piero Gobetti 93/3, 40129 Bologna, Italy\label{aff17}
\and
IFPU, Institute for Fundamental Physics of the Universe, via Beirut 2, 34151 Trieste, Italy\label{aff18}
\and
INAF-Osservatorio Astronomico di Trieste, Via G. B. Tiepolo 11, 34143 Trieste, Italy\label{aff19}
\and
INFN, Sezione di Trieste, Via Valerio 2, 34127 Trieste TS, Italy\label{aff20}
\and
SISSA, International School for Advanced Studies, Via Bonomea 265, 34136 Trieste TS, Italy\label{aff21}
\and
Dipartimento di Fisica e Astronomia, Universit\`a di Bologna, Via Gobetti 93/2, 40129 Bologna, Italy\label{aff22}
\and
INFN-Sezione di Bologna, Viale Berti Pichat 6/2, 40127 Bologna, Italy\label{aff23}
\and
Dipartimento di Fisica, Universit\`a di Genova, Via Dodecaneso 33, 16146, Genova, Italy\label{aff24}
\and
INFN-Sezione di Genova, Via Dodecaneso 33, 16146, Genova, Italy\label{aff25}
\and
Department of Physics "E. Pancini", University Federico II, Via Cinthia 6, 80126, Napoli, Italy\label{aff26}
\and
INAF-Osservatorio Astronomico di Capodimonte, Via Moiariello 16, 80131 Napoli, Italy\label{aff27}
\and
Institute for Astronomy, University of Edinburgh, Royal Observatory, Blackford Hill, Edinburgh EH9 3HJ, UK\label{aff28}
\and
Leiden Observatory, Leiden University, Einsteinweg 55, 2333 CC Leiden, The Netherlands\label{aff29}
\and
INAF-IASF Milano, Via Alfonso Corti 12, 20133 Milano, Italy\label{aff30}
\and
INAF-Osservatorio Astronomico di Roma, Via Frascati 33, 00078 Monteporzio Catone, Italy\label{aff31}
\and
INFN-Sezione di Roma, Piazzale Aldo Moro, 2 - c/o Dipartimento di Fisica, Edificio G. Marconi, 00185 Roma, Italy\label{aff32}
\and
Port d'Informaci\'{o} Cient\'{i}fica, Campus UAB, C. Albareda s/n, 08193 Bellaterra (Barcelona), Spain\label{aff33}
\and
Institute for Theoretical Particle Physics and Cosmology (TTK), RWTH Aachen University, 52056 Aachen, Germany\label{aff34}
\and
Deutsches Zentrum f\"ur Luft- und Raumfahrt e. V. (DLR), Linder H\"ohe, 51147 K\"oln, Germany\label{aff35}
\and
INFN section of Naples, Via Cinthia 6, 80126, Napoli, Italy\label{aff36}
\and
Institute for Astronomy, University of Hawaii, 2680 Woodlawn Drive, Honolulu, HI 96822, USA\label{aff37}
\and
Dipartimento di Fisica e Astronomia "Augusto Righi" - Alma Mater Studiorum Universit\`a di Bologna, Viale Berti Pichat 6/2, 40127 Bologna, Italy\label{aff38}
\and
Instituto de Astrof\'{\i}sica de Canarias, E-38205 La Laguna, Tenerife, Spain\label{aff39}
\and
European Space Agency/ESRIN, Largo Galileo Galilei 1, 00044 Frascati, Roma, Italy\label{aff40}
\and
Universit\'e Claude Bernard Lyon 1, CNRS/IN2P3, IP2I Lyon, UMR 5822, Villeurbanne, F-69100, France\label{aff41}
\and
Institut de Ci\`{e}ncies del Cosmos (ICCUB), Universitat de Barcelona (IEEC-UB), Mart\'{i} i Franqu\`{e}s 1, 08028 Barcelona, Spain\label{aff42}
\and
Instituci\'o Catalana de Recerca i Estudis Avan\c{c}ats (ICREA), Passeig de Llu\'{\i}s Companys 23, 08010 Barcelona, Spain\label{aff43}
\and
Institut de Ciencies de l'Espai (IEEC-CSIC), Campus UAB, Carrer de Can Magrans, s/n Cerdanyola del Vall\'es, 08193 Barcelona, Spain\label{aff44}
\and
UCB Lyon 1, CNRS/IN2P3, IUF, IP2I Lyon, 4 rue Enrico Fermi, 69622 Villeurbanne, France\label{aff45}
\and
Mullard Space Science Laboratory, University College London, Holmbury St Mary, Dorking, Surrey RH5 6NT, UK\label{aff46}
\and
Departamento de F\'isica, Faculdade de Ci\^encias, Universidade de Lisboa, Edif\'icio C8, Campo Grande, PT1749-016 Lisboa, Portugal\label{aff47}
\and
Instituto de Astrof\'isica e Ci\^encias do Espa\c{c}o, Faculdade de Ci\^encias, Universidade de Lisboa, Campo Grande, 1749-016 Lisboa, Portugal\label{aff48}
\and
Department of Astronomy, University of Geneva, ch. d'Ecogia 16, 1290 Versoix, Switzerland\label{aff49}
\and
INFN-Padova, Via Marzolo 8, 35131 Padova, Italy\label{aff50}
\and
INAF-Istituto di Astrofisica e Planetologia Spaziali, via del Fosso del Cavaliere, 100, 00100 Roma, Italy\label{aff51}
\and
Universit\'e Paris-Saclay, Universit\'e Paris Cit\'e, CEA, CNRS, AIM, 91191, Gif-sur-Yvette, France\label{aff52}
\and
INFN-Bologna, Via Irnerio 46, 40126 Bologna, Italy\label{aff53}
\and
School of Physics, HH Wills Physics Laboratory, University of Bristol, Tyndall Avenue, Bristol, BS8 1TL, UK\label{aff54}
\and
University Observatory, LMU Faculty of Physics, Scheinerstr.~1, 81679 Munich, Germany\label{aff55}
\and
Max Planck Institute for Extraterrestrial Physics, Giessenbachstr. 1, 85748 Garching, Germany\label{aff56}
\and
INAF-Osservatorio Astronomico di Padova, Via dell'Osservatorio 5, 35122 Padova, Italy\label{aff57}
\and
Universit\"ats-Sternwarte M\"unchen, Fakult\"at f\"ur Physik, Ludwig-Maximilians-Universit\"at M\"unchen, Scheinerstr.~1, 81679 M\"unchen, Germany\label{aff58}
\and
Institute of Theoretical Astrophysics, University of Oslo, P.O. Box 1029 Blindern, 0315 Oslo, Norway\label{aff59}
\and
Jet Propulsion Laboratory, California Institute of Technology, 4800 Oak Grove Drive, Pasadena, CA, 91109, USA\label{aff60}
\and
Felix Hormuth Engineering, Goethestr. 17, 69181 Leimen, Germany\label{aff61}
\and
Technical University of Denmark, Elektrovej 327, 2800 Kgs. Lyngby, Denmark\label{aff62}
\and
Cosmic Dawn Center (DAWN), Denmark\label{aff63}
\and
Max-Planck-Institut f\"ur Astronomie, K\"onigstuhl 17, 69117 Heidelberg, Germany\label{aff64}
\and
NASA Goddard Space Flight Center, Greenbelt, MD 20771, USA\label{aff65}
\and
Department of Physics and Astronomy, University College London, Gower Street, London WC1E 6BT, UK\label{aff66}
\and
Universit\'e de Gen\`eve, D\'epartement de Physique Th\'eorique and Centre for Astroparticle Physics, 24 quai Ernest-Ansermet, CH-1211 Gen\`eve 4, Switzerland\label{aff67}
\and
Department of Physics, P.O. Box 64, University of Helsinki, 00014 Helsinki, Finland\label{aff68}
\and
Helsinki Institute of Physics, Gustaf H{\"a}llstr{\"o}min katu 2, University of Helsinki, 00014 Helsinki, Finland\label{aff69}
\and
Laboratoire d'etude de l'Univers et des phenomenes eXtremes, Observatoire de Paris, Universit\'e PSL, Sorbonne Universit\'e, CNRS, 92190 Meudon, France\label{aff70}
\and
SKAO, Jodrell Bank, Lower Withington, Macclesfield SK11 9FT, UK\label{aff71}
\and
Centre de Calcul de l'IN2P3/CNRS, 21 avenue Pierre de Coubertin 69627 Villeurbanne Cedex, France\label{aff72}
\and
Dipartimento di Fisica "Aldo Pontremoli", Universit\`a degli Studi di Milano, Via Celoria 16, 20133 Milano, Italy\label{aff73}
\and
INFN-Sezione di Milano, Via Celoria 16, 20133 Milano, Italy\label{aff74}
\and
University of Applied Sciences and Arts of Northwestern Switzerland, School of Computer Science, 5210 Windisch, Switzerland\label{aff75}
\and
Universit\"at Bonn, Argelander-Institut f\"ur Astronomie, Auf dem H\"ugel 71, 53121 Bonn, Germany\label{aff76}
\and
Aix-Marseille Universit\'e, CNRS, CNES, LAM, Marseille, France\label{aff77}
\and
Dipartimento di Fisica e Astronomia "Augusto Righi" - Alma Mater Studiorum Universit\`a di Bologna, via Piero Gobetti 93/2, 40129 Bologna, Italy\label{aff78}
\and
Department of Physics, Institute for Computational Cosmology, Durham University, South Road, Durham, DH1 3LE, UK\label{aff79}
\and
Universit\'e Paris Cit\'e, CNRS, Astroparticule et Cosmologie, 75013 Paris, France\label{aff80}
\and
CNRS-UCB International Research Laboratory, Centre Pierre Bin\'etruy, IRL2007, CPB-IN2P3, Berkeley, USA\label{aff81}
\and
Institut d'Astrophysique de Paris, 98bis Boulevard Arago, 75014, Paris, France\label{aff82}
\and
Institut d'Astrophysique de Paris, UMR 7095, CNRS, and Sorbonne Universit\'e, 98 bis boulevard Arago, 75014 Paris, France\label{aff83}
\and
Institute of Physics, Laboratory of Astrophysics, Ecole Polytechnique F\'ed\'erale de Lausanne (EPFL), Observatoire de Sauverny, 1290 Versoix, Switzerland\label{aff84}
\and
Telespazio UK S.L. for European Space Agency (ESA), Camino bajo del Castillo, s/n, Urbanizacion Villafranca del Castillo, Villanueva de la Ca\~nada, 28692 Madrid, Spain\label{aff85}
\and
Institut de F\'{i}sica d'Altes Energies (IFAE), The Barcelona Institute of Science and Technology, Campus UAB, 08193 Bellaterra (Barcelona), Spain\label{aff86}
\and
European Space Agency/ESTEC, Keplerlaan 1, 2201 AZ Noordwijk, The Netherlands\label{aff87}
\and
School of Mathematics, Statistics and Physics, Newcastle University, Herschel Building, Newcastle-upon-Tyne, NE1 7RU, UK\label{aff88}
\and
DARK, Niels Bohr Institute, University of Copenhagen, Jagtvej 155, 2200 Copenhagen, Denmark\label{aff89}
\and
Waterloo Centre for Astrophysics, University of Waterloo, Waterloo, Ontario N2L 3G1, Canada\label{aff90}
\and
Department of Physics and Astronomy, University of Waterloo, Waterloo, Ontario N2L 3G1, Canada\label{aff91}
\and
Perimeter Institute for Theoretical Physics, Waterloo, Ontario N2L 2Y5, Canada\label{aff92}
\and
Space Science Data Center, Italian Space Agency, via del Politecnico snc, 00133 Roma, Italy\label{aff93}
\and
Centre National d'Etudes Spatiales -- Centre spatial de Toulouse, 18 avenue Edouard Belin, 31401 Toulouse Cedex 9, France\label{aff94}
\and
Institute of Space Science, Str. Atomistilor, nr. 409 M\u{a}gurele, Ilfov, 077125, Romania\label{aff95}
\and
Dipartimento di Fisica e Astronomia "G. Galilei", Universit\`a di Padova, Via Marzolo 8, 35131 Padova, Italy\label{aff96}
\and
Institut f\"ur Theoretische Physik, University of Heidelberg, Philosophenweg 16, 69120 Heidelberg, Germany\label{aff97}
\and
Universit\'e St Joseph; Faculty of Sciences, Beirut, Lebanon\label{aff98}
\and
Departamento de F\'isica, FCFM, Universidad de Chile, Blanco Encalada 2008, Santiago, Chile\label{aff99}
\and
Universit\"at Innsbruck, Institut f\"ur Astro- und Teilchenphysik, Technikerstr. 25/8, 6020 Innsbruck, Austria\label{aff100}
\and
Department of Physics and Helsinki Institute of Physics, Gustaf H\"allstr\"omin katu 2, University of Helsinki, 00014 Helsinki, Finland\label{aff101}
\and
Department of Physics, Royal Holloway, University of London, Surrey TW20 0EX, UK\label{aff102}
\and
Instituto de Astrof\'isica e Ci\^encias do Espa\c{c}o, Faculdade de Ci\^encias, Universidade de Lisboa, Tapada da Ajuda, 1349-018 Lisboa, Portugal\label{aff103}
\and
Cosmic Dawn Center (DAWN)\label{aff104}
\and
Niels Bohr Institute, University of Copenhagen, Jagtvej 128, 2200 Copenhagen, Denmark\label{aff105}
\and
Universidad Polit\'ecnica de Cartagena, Departamento de Electr\'onica y Tecnolog\'ia de Computadoras,  Plaza del Hospital 1, 30202 Cartagena, Spain\label{aff106}
\and
Kapteyn Astronomical Institute, University of Groningen, PO Box 800, 9700 AV Groningen, The Netherlands\label{aff107}
\and
Caltech/IPAC, 1200 E. California Blvd., Pasadena, CA 91125, USA\label{aff108}
\and
Dipartimento di Fisica e Scienze della Terra, Universit\`a degli Studi di Ferrara, Via Giuseppe Saragat 1, 44122 Ferrara, Italy\label{aff109}
\and
Istituto Nazionale di Fisica Nucleare, Sezione di Ferrara, Via Giuseppe Saragat 1, 44122 Ferrara, Italy\label{aff110}
\and
INAF, Istituto di Radioastronomia, Via Piero Gobetti 101, 40129 Bologna, Italy\label{aff111}
\and
Astronomical Observatory of the Autonomous Region of the Aosta Valley (OAVdA), Loc. Lignan 39, I-11020, Nus (Aosta Valley), Italy\label{aff112}
\and
Universit\'e C\^{o}te d'Azur, Observatoire de la C\^{o}te d'Azur, CNRS, Laboratoire Lagrange, Bd de l'Observatoire, CS 34229, 06304 Nice cedex 4, France\label{aff113}
\and
ICSC - Centro Nazionale di Ricerca in High Performance Computing, Big Data e Quantum Computing, Via Magnanelli 2, Bologna, Italy\label{aff114}
\and
Univ. Grenoble Alpes, CNRS, Grenoble INP, LPSC-IN2P3, 53, Avenue des Martyrs, 38000, Grenoble, France\label{aff115}
\and
Dipartimento di Fisica, Sapienza Universit\`a di Roma, Piazzale Aldo Moro 2, 00185 Roma, Italy\label{aff116}
\and
Aurora Technology for European Space Agency (ESA), Camino bajo del Castillo, s/n, Urbanizacion Villafranca del Castillo, Villanueva de la Ca\~nada, 28692 Madrid, Spain\label{aff117}
\and
Dipartimento di Fisica - Sezione di Astronomia, Universit\`a di Trieste, Via Tiepolo 11, 34131 Trieste, Italy\label{aff118}
\and
Department of Mathematics and Physics E. De Giorgi, University of Salento, Via per Arnesano, CP-I93, 73100, Lecce, Italy\label{aff119}
\and
INFN, Sezione di Lecce, Via per Arnesano, CP-193, 73100, Lecce, Italy\label{aff120}
\and
INAF-Sezione di Lecce, c/o Dipartimento Matematica e Fisica, Via per Arnesano, 73100, Lecce, Italy\label{aff121}
\and
ICL, Junia, Universit\'e Catholique de Lille, LITL, 59000 Lille, France\label{aff122}
\and
Instituto de F\'isica Te\'orica UAM-CSIC, Campus de Cantoblanco, 28049 Madrid, Spain\label{aff123}
\and
CERCA/ISO, Department of Physics, Case Western Reserve University, 10900 Euclid Avenue, Cleveland, OH 44106, USA\label{aff124}
\and
Laboratoire Univers et Th\'eorie, Observatoire de Paris, Universit\'e PSL, Universit\'e Paris Cit\'e, CNRS, 92190 Meudon, France\label{aff125}
\and
Departamento de F{\'\i}sica Fundamental. Universidad de Salamanca. Plaza de la Merced s/n. 37008 Salamanca, Spain\label{aff126}
\and
Universit\'e de Strasbourg, CNRS, Observatoire astronomique de Strasbourg, UMR 7550, 67000 Strasbourg, France\label{aff127}
\and
Center for Data-Driven Discovery, Kavli IPMU (WPI), UTIAS, The University of Tokyo, Kashiwa, Chiba 277-8583, Japan\label{aff128}
\and
Jodrell Bank Centre for Astrophysics, Department of Physics and Astronomy, University of Manchester, Oxford Road, Manchester M13 9PL, UK\label{aff129}
\and
California Institute of Technology, 1200 E California Blvd, Pasadena, CA 91125, USA\label{aff130}
\and
Department of Physics \& Astronomy, University of California Irvine, Irvine CA 92697, USA\label{aff131}
\and
Departamento F\'isica Aplicada, Universidad Polit\'ecnica de Cartagena, Campus Muralla del Mar, 30202 Cartagena, Murcia, Spain\label{aff132}
\and
Instituto de F\'isica de Cantabria, Edificio Juan Jord\'a, Avenida de los Castros, 39005 Santander, Spain\label{aff133}
\and
Institute of Cosmology and Gravitation, University of Portsmouth, Portsmouth PO1 3FX, UK\label{aff134}
\and
Department of Computer Science, Aalto University, PO Box 15400, Espoo, FI-00 076, Finland\label{aff135}
\and
Universidad de La Laguna, Dpto. Astrof\'\i sica, E-38206 La Laguna, Tenerife, Spain\label{aff136}
\and
Department of Physics and Astronomy, Vesilinnantie 5, University of Turku, 20014 Turku, Finland\label{aff137}
\and
Finnish Centre for Astronomy with ESO (FINCA), Quantum, Vesilinnantie 5, University of Turku, 20014 Turku, Finland\label{aff138}
\and
Serco for European Space Agency (ESA), Camino bajo del Castillo, s/n, Urbanizacion Villafranca del Castillo, Villanueva de la Ca\~nada, 28692 Madrid, Spain\label{aff139}
\and
ARC Centre of Excellence for Dark Matter Particle Physics, Melbourne, Australia\label{aff140}
\and
Centre for Astrophysics \& Supercomputing, Swinburne University of Technology,  Hawthorn, Victoria 3122, Australia\label{aff141}
\and
Department of Physics, Oxford University, Keble Road, Oxford OX1 3RH, UK\label{aff142}
\and
DAMTP, Centre for Mathematical Sciences, Wilberforce Road, Cambridge CB3 0WA, UK\label{aff143}
\and
Kavli Institute for Cosmology Cambridge, Madingley Road, Cambridge, CB3 0HA, UK\label{aff144}
\and
Department of Astrophysics, University of Zurich, Winterthurerstrasse 190, 8057 Zurich, Switzerland\label{aff145}
\and
Department of Physics, Centre for Extragalactic Astronomy, Durham University, South Road, Durham, DH1 3LE, UK\label{aff146}
\and
IRFU, CEA, Universit\'e Paris-Saclay 91191 Gif-sur-Yvette Cedex, France\label{aff147}
\and
INAF-Osservatorio Astrofisico di Arcetri, Largo E. Fermi 5, 50125, Firenze, Italy\label{aff148}
\and
Centro de Astrof\'{\i}sica da Universidade do Porto, Rua das Estrelas, 4150-762 Porto, Portugal\label{aff149}
\and
Dipartimento di Fisica, Universit\`a di Roma Tor Vergata, Via della Ricerca Scientifica 1, Roma, Italy\label{aff150}
\and
INFN, Sezione di Roma 2, Via della Ricerca Scientifica 1, Roma, Italy\label{aff151}
\and
HE Space for European Space Agency (ESA), Camino bajo del Castillo, s/n, Urbanizacion Villafranca del Castillo, Villanueva de la Ca\~nada, 28692 Madrid, Spain\label{aff152}
\and
INAF - Osservatorio Astronomico d'Abruzzo, Via Maggini, 64100, Teramo, Italy\label{aff153}
\and
Mathematical Institute, University of Leiden, Einsteinweg 55, 2333 CA Leiden, The Netherlands\label{aff154}
\and
Institute of Astronomy, University of Cambridge, Madingley Road, Cambridge CB3 0HA, UK\label{aff155}
\and
Center for Astrophysics and Cosmology, University of Nova Gorica, Nova Gorica, Slovenia\label{aff156}
\and
Institute for Particle Physics and Astrophysics, Dept. of Physics, ETH Zurich, Wolfgang-Pauli-Strasse 27, 8093 Zurich, Switzerland\label{aff157}
\and
Department of Astrophysical Sciences, Peyton Hall, Princeton University, Princeton, NJ 08544, USA\label{aff158}
\and
Space physics and astronomy research unit, University of Oulu, Pentti Kaiteran katu 1, FI-90014 Oulu, Finland\label{aff159}
\and
Institut de Physique Th\'eorique, CEA, CNRS, Universit\'e Paris-Saclay 91191 Gif-sur-Yvette Cedex, France\label{aff160}
\and
International Centre for Theoretical Physics (ICTP), Strada Costiera 11, 34151 Trieste, Italy\label{aff161}
\and
Center for Computational Astrophysics, Flatiron Institute, 162 5th Avenue, 10010, New York, NY, USA\label{aff162}}    

%
%
\abstract{
One of the \Euclid mission's key projects is the so-called 3$\times$2pt analysis, that is,  the combination of cosmic shear, photometric galaxy clustering, and galaxy-galaxy lensing. 
Although \Euclid has established quality requirements for the photo-$z$ accuracy needed for the weak lensing galaxy sample, 
no such requirements have been set for the photometric clustering sample. 
In this paper, we investigate the impact of redshift uncertainties on \Euclid's photometric galaxy clustering analysis and its combination with weak gravitational lensing, focusing on data release 1 (DR1). In particular, we study whether having precise knowledge of the mean of the redshift distributions per bin is sufficient to avoid biases in the resulting cosmological constraints or whether accuracy in the higher-order moments of the distribution is required. We evaluate the results based on their constraining power on $w_{\mathrm{0}}$ and $w_{a}$ and define thresholds for the precision and accuracy of \Euclid's redshift distribution of the photometric clustering sample. 
We find that the redshift distributions of the photometric clustering sample must be known at an accuracy of 0.004(1+$z$) in  the mean in order to recover 80$\%$ of the constraining power in \Euclid's DR1 $w_0w_{a}$CDM 3$\times$2pt analysis.
The impact of the uncertainty on the width is negligible, provided the mean redshift is constrained with sufficient accuracy. For most sources of redshift distribution error, attaining the requirement on the mean will also reduce uncertainty in the width well below the required level.
}
%
%
%
%
   \titlerunning{Impact of redshift distribution uncertainties}
   \authorrunning{Euclid Collaboration: Bertmann et al.}
   
   \maketitle
%
%
%
%
   
\section{\label{sc:Intro}Introduction}
The ongoing \Euclid mission of the European Space Agency is a cornerstone of European space exploration.
Since the satellite's launch on 1 July 2023 it probes the distribution and nature of matter in the Universe across large scales, up to redshifts of approximately $z\,=\,2$.
The survey is specifically designed for the investigation of dark matter and dark energy using a combination of optical and near-infrared imaging as well as spectroscopy \citep{Laureijs11, EuclidSkyOverview}.
An essential component of the mission is the so-called 3$\times$2pt analysis, the combination of galaxy clustering, 
weak gravitational lensing, and their cross-correlations also known as galaxy-galaxy lensing.
Galaxy clustering -- the autocorrelation of galaxy positions -- serves as a primary tool to investigate the underlying matter density field and its fluctuations, though its interpretation requires accounting for galaxy bias, which describes how gala\-xies statistically trace the mass distribution. 
The weak len\-sing signal inferred from source galaxy shapes is sensitive to the total amount of matter along the line of sight and thus to the amplitude of matter fluctuations and gravity \citep{Bartelmann2001}.
The combination of these complementary probes is of great relevance to break the \Omm--\,$\sigma_{\rm 8}$ degeneracy that arises in shear-only analyses, as it has been demonstrated in previous analyses from the Dark Energy Survey (DES, \citealt{abbott2022dark}), the Kilo-Degree Survey (KiDS, \citealt{heymans2021kids}), and the Hyper Suprime-Cam Subaru Strategic Program (HSC-SSP, \citealt{sugiyama2023hyper,miyatake2023hsc3x2}). 

The 3$\times$2pt analysis requires two galaxy samples: 
the so-called lens sample, used to measure photometric galaxy clustering, and the source sample, used to measure the weak gra\-vitational lensing signal. While using a single galaxy sample for both sources and lenses has been explored in recent literature (see, e.g., \citealt{2020Schan,2024Boruah,2025Xiong}), most data ana\-ly\-ses use two different sample selections to optimise the signal-to-noise ratio of photometric galaxy clustering and weak lensing, respectively. Yet, having two different samples (source and lens) does not necessarily imply that their galaxy overlap is zero. 
The source sample consists of distant galaxies whose light is distorted by the gravitational lensing effect when travelling across the large-scale structure between those galaxies and the observer. 
The lens sample is used to measure both photometric galaxy clustering and galaxy-galaxy len\-sing. Therefore, in the latter, 
the lens galaxies and their underlying mass distribution act as gravitational lenses, bending the light emitted by the source galaxies and other background objects.
Accurate distance measurements between source and lens galaxies critically depend on comprehensive redshift information for both. 
These redshift measurements are fundamental to reconstruct the 3D distribution of matter in the Universe, and are used to constrain cosmological parameters, such as the amount of dark matter and dark energy.
Most of the redshifts measured by \Euclid are photometric redshifts (photo-$z$s), since \Euclid's slitless spectrometer has a limiting flux of $2\times10^{-16}\;\mathrm{erg\,s^{-1}\,cm^{-2}}$ and a restricted range of spectroscopic redshifts $0.9<z<1.8$.
The Euclid Collaboration has established specific quality standards for the accuracy of photo-$z$s measured for the source galaxies \citep{EuclidSkyOverview}.
However, similar standards for the photo-$z$s of the lens galaxies, which are used to study galaxy clustering and galaxy-galaxy lensing, have not been explored yet.
The entire galaxy sample is typically divided into several redshift bins. 
Knowledge of the precise shape of the redshift distribution $n(z)$ in each bin is essential for the success of the subsequent analysis.
Therefore, it is of critical importance to consider the uncertainty in the redshift distributions when deriving cosmological constraints. 
For cosmic shear (see \citealt{kilbinger2015cosmology} for a review), the mean of the redshift distribution in each bin is a main source of uncertainty affecting the inferred constraints. 
However, in photometric galaxy clustering analyses, the strength of the clustering signal is directly influenced by, and thus critically sensitive to, the shape or width of the lens redshift distribution \citep{stretch}.

In this work we examine how uncertainties in both the mean and width of the lens redshift distributions can impact the cosmological constraints obtained from the \Euclid data. 
Our main analysis consists of a forecast with simulated likelihoods using Monte Carlo methods on noiseless theory data vectors. 
Additionally, we test the results against Fisher forecasts on a similar setup and also consider measurements from the Flagship simulation \citep{EuclidSkyFlagship}.
In this work we focus on the upcoming data release 1 (DR1), which will include the first year of \Euclid observations.
However, the impact of such redshift uncertainties is generally applicable to other scenarios as well.
For instance, \citet{Blot25} and \citet{canas2025euclid} study a variety of systematic effects, including uncertainties in the mean of the photometric redshift distributions, in the context of data release 3 (DR3). 
Another recent study of uncertainties in the photo-$z$ distribution of the sources is presented in \citet{lsstforecast} against the backdrop of the upcoming Legacy Survey of Space and Time \citep[LSST,][]{LSST}.
\\

This paper is structured as follows: Section \ref{sc:theory} details the modelling of the 2-point correlation functions of which the data vectors are composed.
In Sect. \ref{sc:flagship}, we review the Flagship simulation
and the inputs taken from it for the analysis.
Our forecasting methodology is presented in Sect. \ref{sc:Method}. 
The results of our analysis are shown in Sect. \ref{sc:Results}, and a conclusion is given in Sect. \ref{sc:Conclusions}.

\section{\label{sc:theory}Theoretical modelling}

A key aspect of \Euclid's mission is the 3$\times$2pt analysis, namely the combination of the individual probes of galaxy clustering, galaxy-galaxy lensing, and cosmic shear. 
These probes are the auto- and cross-correlations of lens galaxy positions and source galaxy shapes.

\subsection{\label{subs:clustering}Galaxy clustering}
Galaxies are not distributed uniformly in the Universe but rather tend to cluster together on various scales, ranging from gravitationally bound groups and galaxy clusters to larger, more loosely associated structures.
We can measure the excess (over an expected random distribution) number of galaxy pairs separated by an angular distance $\theta$, $w(\theta)$.
Galaxy clustering can be used as a probe of the growth of structure of the Universe. However, one of its primary sources of uncertainty is the so-called galaxy bias that arises from the fact that galaxies do not trace the exact underlying distribution of the dark matter.
This leads to a difference between the matter power spectrum and the galaxy power spectrum \citep{bardeen1986statistics, kaiser1987clustering,1993Fry&Gaztanaga}.
In the linear approximation and in real space, this is described by a bias factor
\begin{equation}
    P_{\mathrm{g}}(k) = b^2 P(k) \, ,
\end{equation}
with the matter power spectrum $P(k)$ and the galaxy power spectrum $P_{\mathrm{g}}(k)$.

We neglect here the possible scale dependence of the bias, as well as non-linearity and stochasticity.
This approximation is valid on large scales (see, e.g., \citealt{abbott2018dark}).
The galaxy clustering angular correlation function can be written as 
\begin{equation}\label{eq:w_theta}
    w^{i}(\theta) = \sum_{\ell\geq0} \frac{2\ell +1}{4\pi} d_{00}^\ell(\theta) \, C^{ii}_{\mathrm{\delta_{\mathrm{g}}}\mathrm{\delta_{\mathrm{g}}}}\!\left(\ell\right) \; ,
\end{equation}
where \(d_{mn}^\ell\) is the Wigner $d$-matrix, such that \(d_{00}^\ell(\theta)=P_{\ell}\left(\cos{\theta}\right)\).
The expression includes the angular clustering power spectrum
\begin{equation}  
\begin{split}
C^{ii}_{\mathrm{\delta_{\mathrm{g}}}\mathrm{\delta_{\mathrm{g}}}}\left(\ell\right) = &\frac{2}{\pi} \int_{\chi_{\rm1,min}}^{\chi_{\rm1,max}} \mathrm{d}\chi_1 \, q_{\mathrm{\delta_{\mathrm{g}}}}^{i}\!\left(\chi_1\right) \int_{\chi_{\rm2,min}}^{\chi_{\rm2,max}} \mathrm{d}\chi_2 \, q_{\mathrm{\delta_{\mathrm{g}}}}^{i}\!\left(\chi_2\right)\\ &\int_0^{\infty} \frac{\mathrm{d}k}{k} k^3 
P_{\mathrm{lin}}\left(k\right)\, T\left(k,\chi_1\right)\, T\left(k,\chi_2\right) j_{\ell}\left(k\chi_1\right) j_{\ell}\left(k\chi_2\right)\, ,
\end{split}
\label{eq:C_ells}
\end{equation}
with the $\ell$-th-order spherical Bessel function $j_{\ell}$, the linear matter power spectrum $P_{\mathrm{lin}}(k)$, 
the transfer function of matter perturbations $T(k, \chi_i)$, and the density kernel
\begin{equation}
    q_{\mathrm{\delta_{\mathrm{g}}}}^{i}(\chi) = b^{i}\left[z(\chi)\right] n_{\mathrm{g}}^{i}\left[z(\chi)\right]\frac{\mathrm{d}z}{\mathrm{d}\chi}\,, 
\end{equation}
with the normalised redshift distribution of lens galaxies in the $i$-th bin $n_{\mathrm{g}}^{i}$. 
The integral limits $\chi_{\rm min}$ and  $\chi_{\rm max}$ correspond to the values of the comoving distance at the minimum and maximum redshifts of the redshift distribution $n_{\mathrm{g}}^{i}(z)$ in each bin. 
Although not accounted for in Eq.~\eqref{eq:C_ells}, we do include the effect of redshift space distortions and lens magnification in the computation of the angular clustering power spectrum $C^{ii}_{\mathrm{\delta_{\mathrm{g}}}\mathrm{\delta_{\mathrm{g}}}}\!\left(\ell\right)$. The full expression including those terms is described in \cite{Fang_2020}.

\subsection{\label{subs:gglensing}Galaxy-galaxy lensing}
The lensing of a background galaxy by the mass distribution of foreground galaxies is referred to as galaxy-galaxy lensing.
Therefore, it is the cross-correlation of lens position with source ellipticity.
Characteristic for this effect is a distortion in the form of a tangential shear $\gamma_{\mathrm{t}}(\theta)$ that depends on the mass of the foreground object.
It can be constructed similarly to the galaxy clustering angular correlation function by correlating the lens galaxy positions in the $i$-th bin with the source galaxy shear in the $j$-th bin:
\begin{equation}
    \begin{split}
    \gamma_{\mathrm{t}}^{ij} (\theta) = 
    \sum_{\ell\geq2} \frac{2\ell +1}{4\pi} C^{ij}_{\mathrm{\delta_{\mathrm{g}}}\mathrm{\kappa}}\!\left(\ell\right)\, d_{02}^\ell\!\left(\theta\right)  \; .
    \end{split}
\end{equation}
The relation above introduces the angular galaxy-galaxy lensing power spectrum, $C^{ij}_{\mathrm{\delta_{\mathrm{g}}}\mathrm{\kappa}}\left(\ell\right)$. Moreover, we apply the Limber approximation \citep{Limber1953, loverde2008extended}, yielding
\begin{equation}  
\begin{split}
C^{ij}_{\mathrm{\delta_{\mathrm{g}}}\mathrm{\kappa}}\!\left(\ell\right) = \int_{\mathrm{0}}^{\mathrm{\chi_{\mathrm{H}}}} \mathrm{d}\chi \frac{q_{\mathrm{\delta_{\mathrm{g}}}}^{i}\!\left(\chi\right) \, q_{\mathrm{\mathrm{\kappa}}}^{j}(\chi)}{\chi^2} \; P\left(\frac{\ell+\frac{1}{2}}{\chi}, \chi\right)\;,
\end{split}
\end{equation}
with the power spectrum $P(k,\chi)$, the Hubble horizon $\chi_{\mathrm{H}}$, and the lensing efficiency for a spatially flat universe
\begin{equation}\label{lensingkernel}
    q_{\mathrm{\mathrm{\kappa}}}^{i}(\chi) = \frac{3 H_{\mathrm{0}}^2\,\Omega_{\mathrm{m}}}{2c^2}\frac{\chi}{a(\chi)} \int_{\chi}^{\chi_{\mathrm{H}}}  \mathrm{d}\chi' n_{\mathrm{\kappa}}^{i}\left[z(\chi')\right]\,\frac{\mathrm{d}z}{\mathrm{d}\chi'} \frac{\chi'-\chi}{\chi'} \, .
\end{equation}

Here, $H_{\mathrm{0}}$ denotes the Hubble constant, $\Omega_{\mathrm{m}}$ the matter density parameter, $c$ the speed of light, $a$ the scale factor, and $n_{\mathrm{\kappa}}^{i}(z)$ 
is the normalised redshift distribution of the source galaxy sample.
Given that the tangential shear $\gamma_{\mathrm{t}}$ is a non-local measure of the galaxy-mass cross-correlation, we adopt analytic marginalisation over the mass enclosed below the angular scales included in the analysis (point-mass marginalisation) to mitigate this effect. Since $\gamma_{\rm t}(\theta)$ depends on the integrated mass inside $\theta$, 
uncertainties on small (excluded) scales would otherwise propagate to the larger scales used in the analysis, 
making standard scale cuts insufficient. We apply this marginalisation to all scales considered in our analysis 
(for details on the scale cuts see Sect. \ref{subs scalecuts}), ensuring that baryonic effects and non-linearities 
from the inner regions are effectively removed. We refer the reader to \cite{MacCrann_2020} and \cite{Pandey_2022} for a detailed description of the implementation and validation of this approach.

\subsection{\label{subs:shear}Cosmic shear}
The term cosmic shear describes the weak lensing effect induced by the foreground large-scale structure and is the auto-correlation of source galaxy shapes.
The shear-shear 2-point functions are typically expressed as the sum or difference of the products of the tangential and cross components of the shear after averaging over all galaxy pairs at a given separation $\theta$,
\begin{equation}
    \xi_{+}(\theta) = \langle \gamma_{\mathrm{t}}\gamma_{\mathrm{t}}\rangle (\theta) + \langle \gamma_{\mathrm{\times}}\gamma_{\mathrm{\times}}\rangle (\theta) \, , \; \xi_{-}(\theta) = \langle \gamma_{\mathrm{t}}\gamma_{\mathrm{t}}\rangle (\theta) - \langle \gamma_{\mathrm{\times}}\gamma_{\mathrm{\times}}\rangle (\theta) \, .
\end{equation}

Applied to tomography and assuming the Limber approximation, this yields the correlation functions
\begin{equation}
    \begin{split}
    \xi_{+/-}^{ij}(\theta) = &\sum_{\ell\geq2} \frac{2\ell +1}{4\pi} \left[ C^{ij}_{EE}\!\left(\ell\right) \pm C^{ij}_{BB}\!\left(\ell\right) \right]\, d_{\pm 2\,2}^\ell\!\left(\theta\right) \\ + &\sum_{\ell\geq2} \frac{2\ell +1}{4\pi}\, i \left[ C^{ij}_{BE}\!\left(\ell\right) \mp C^{ij}_{EB}\!\left(\ell\right) \right]\, d_{\pm2\,2}^\ell\!\left(\theta\right) \; .
    \end{split}
\end{equation}
as described in \citet{cardone2025cosmology} and with
\begin{equation}  
\begin{split}
C^{ij}_{EE/BB}\!\left(\ell\right) = \int_{\mathrm{0}}^{\mathrm{\chi_{\mathrm{H}}}} \mathrm{d}\chi \frac{q_{\mathrm{\kappa}}^{i}\!\left(\chi\right) \, q_{\mathrm{\mathrm{\kappa}}}^{j}(\chi)}{\chi^2} \; P_{E/B}\left(\frac{\ell+\frac{1}{2}}{\chi}, \chi\right)\;,
\end{split}
\end{equation}
for redshift bins $i$ and $j$ with the respective lensing efficiencies (see Eq. \ref{lensingkernel}) and Bessel functions and the $E$- and $B$-mode power spectra $P_{E/B}\,(k, \chi)$.
We assume the Universe to be parity-invariant, which implies that the expectation values of the cross-correlation spectra $C^{ij}_{\mathrm{B}\mathrm{E}}\!\left(\ell\right)$ and $C^{ij}_{\mathrm{E}\mathrm{B}}\!\left(\ell\right)$ vanish.
For the cosmic shear analysis, we mitigate the impact of small-scale modelling uncertainties such as baryonic feedback by imposing strict angular scale cuts, rather than point-mass marginalisation. We detail the specific scale choices and their 
motivation in Sect. \ref{subs scalecuts}.

\subsection{\label{subs_photoz}Photometric redshifts}

Most redshifts measured by \Euclid will be photo-$z$s.
In contrast to the more accurate but also more limited (in terms of flux and redshift range) technique of spectroscopic measurements, photometry relies on the use of broad filters to measure an object's brightness at different wavelengths (see \citealt{newman2022photometric} for a recent review).
While remarkably efficient for large-scale surveys, photo-$z$s come with inherent limitations compared to their spectroscopic counterparts. 
These limitations are typically characterised using three key properties: bias, scatter, and outlier fraction. 
Understanding and mitigating these properties will be crucial for the scientific goals of missions like \Euclid.
\begin{itemize}
    \item Photo-$z$ bias: 
    the systematic offset between estimated photo-$z$ and true spectroscopic redshift. Sources of bias can include incomplete or inaccurate galaxy spectral energy distribution (SED) templates used in fitting methods, mismatches between the observed filter set and the intrinsic galaxy colours, or even issues with the photometric calibration. Controlling bias is essential for cosmological analyses, as it can systematically shift derived cosmological parameters. 
    \item Photo-$z$ scatter: this term 
    refers to the random uncertainty, or spread, in the photo-$z$ estimates around the true redshift. Scatter arises from factors such as photometric noise, degeneracies in the galaxy colour space (where different combinations of galaxy type and redshift can produce similar observed colours), and the inherent broadband nature of the photometric filters, which provides less spectral information than a full spectrum.
    \item Outlier fraction: this represents the percentage of galaxies for which the photo-$z$ estimate is catastrophically wrong. Outliers typically occur when the photometric data is highly ambiguous, perhaps due to unusual galaxy types not well-represented in the SED templates, or significant blending with other sources in crowded fields. Even a small outlier fraction can significantly impact cosmological analyses, particularly those sensitive to tails of the redshift distribution or requiring highly pure samples.
\end{itemize}

\paragraph{Uncertainties in $n(z)$:} the accuracy of the photo-$z$ distributions is essential for a successful analysis of galaxy clustering and weak lensing.
Whereas this work focuses on the impact of photometric uncertainties of the lens distribution, we also consider, in some of the scenarios, uncertainties in the mean of the source redshift distribution, to make the forecasts more realistic. 
The systematic uncertainties on the mean of each redshift bin are parametrised via a coherent shift per bin
\begin{equation}\label{meanuncertainty}
    n_{\mathrm{true}}^{i}\,=\,n^{i}\;(z-\Delta z^{i})
\end{equation}
to account for the effect of unknown biases.
Uncertainties in the width and therefore the shape of the redshift distributions are described by a stretch $\sigma_z$ of the distribution of the form
\begin{equation}\label{widthuncertainty}
    n_{\mathrm{true}}^{i}(z)\,=\,\frac{1}{\sigma_z^{i}}\,n^{i}\left(\frac{z-z_{\mathrm{mean}}^{i}}{\sigma_z^{i}}+z_{\mathrm{mean}}^{i}\right)\,.
\end{equation}

Here, $z_{\mathrm{mean}}^{i}$ represents the mean redshift of an individual tomographic 
bin. While these equations describe the isolated effects of shifting and stretching the redshift distributions, our analysis also includes cases where both parameters are varied simultaneously.

We note here that we did not include the effect of outliers, tails, or other modes of the redshift distributions and used well-behaved redshift distributions. Future work could consider a more complex approach (see, e.g.,  \citealt{bernstein2025sampling}, \citealt{giannini2025dark}, \citealt{yin2025dark}, and \citealt{d2025dark}, where the uncertainties on the most important modes of the $n(z)$ are considered), even though our work and other investigations (see \citealt{reischke2024propagating}) have shown that the uncertainty in the mean and width dominate the requirements compared to higher-order moments.

\subsection{\label{subs:systematics}Systematic effects}
In this analysis, several systematic effects are taken into account.
These arise mainly from uncertainties in photometric measurements or astrophysical effects that are difficult to measure or model. 
We do not include effects such as foreground removal or image processing. 
For a more detailed discussion of observational effects see \cite{instrumentsys2} and \cite{Paykari-EP6}.
We give here a brief description of the astrophysical and observational effects included in this work and their parametrisation using additional nuisance parameters.\\ 

\paragraph{Multiplicative shear bias:} multiplicative shear bias is a systematic effect in the measurement of the source galaxy shapes.
It therefore affects the galaxy-galaxy lensing as well as the cosmic shear signal.
This bias is caused by se\-veral effects, including mismatches in the model assumptions, point spread function modelling, or selection effects \citep{massey2013origins, hoekstra2017study}.
This can potentially lead to a significant shift of the overall amplitude of the shear measurement and consequently the correlations.
We account for this using a parameter $m^i$ in each source redshift bin \citep{kitching2019propagating} via 
\begin{align}
    \gamma_{\mathrm{t}}^{ij} (\theta) &\longrightarrow (1 + m^{j}) \, \gamma_{\mathrm{t}}^{ij} (\theta) \, , \nonumber\\
    \xi_{+/-}^{ij}(\theta) &\longrightarrow (1 + m^{i})\,(1 + m^{j})\, \xi_{+/-}^{ij}(\theta) \, .
\end{align}
Moreover, we assume that additive shear bias can be accounted for empirically (\citealt{hoekstra2021accounting, li2023kids}).

\paragraph{Magnification bias:} magnification bias is a systematic effect caused by the fact that gravitational lensing not only distorts the shapes of distant galaxies -- resulting in shear -- but also magnifies their observed brightness and changes their apparent number density on the sky.
This magnification effect can pull intrinsically faint galaxies above the survey's detection threshold, increasing the observed number of galaxies, while simultaneously spreading out the sky area of background sources, causing a reduction in number density. 
The magnification bias parameter $\alpha$ is typically measured from the observed luminosity function of the galaxy sample used and indicates the evolution of the power law index from the faint to the bright end of a survey. 
Therefore, it quantifies the sensitivity of the observed density contrast to changes in flux/magnitude due to lensing and causes changes in the observed angular power spectra:
\begin{align}
C^{ii}_{\mathrm{\delta_{\mathrm{g}}}\mathrm{\delta_{\mathrm{g}}}}\!\left(\ell\right) &\longrightarrow  \, C^{ii}_{\mathrm{\delta_{\mathrm{g}}}\mathrm{\delta_{\mathrm{g}}}}\left(\ell\right) + C^{ii}_{\mathrm{\delta_{\mathrm{g}}}\mathrm{M}}\left(\ell\right) + C^{ii}_{\mathrm{M}\mathrm{\delta_{\mathrm{g}}}}\!\left(\ell\right) + C^{ii}_{\mathrm{MM}}\left(\ell\right) \, , \nonumber\\
    C^{ij}_{\mathrm{\delta_{\mathrm{g}}}\mathrm{\kappa}}\!\left(\ell\right) &\longrightarrow \, C^{ij}_{\mathrm{\delta_{\mathrm{g}}}\mathrm{\kappa}}\left(\ell\right) + C^{ij}_{\mathrm{M}\mathrm{\kappa}}\left(\ell\right)  \, .
\end{align}
The details are described in \citet{thiele2020disentangling}. 
The values for the lens magnification parameters are taken from the Flagship simulation \citep{EuclidSkyFlagship}.\\

\paragraph{Intrinsic alignments:} the measured shear of the source galaxies is not only caused by the gravitational field along the line of sight.
Tidal fields, as well as the larger-scale gravitational fields, 
lead to a systematic uncertainty in the weak lensing correlations.
Intrinsic alignments (IA, see \citealt{joachimi2015galaxy} for a review) affect the cosmic shear and galaxy-galaxy lensing signals. 
The IA signal is a particularly large fractional contaminant in the lower source redshift bins because the gravitational lensing signal is inherently weaker at low redshift, making the IA contribution relatively larger than the true cosmological signal it is contaminating.
It is often a challenge to disentangle the two effects, as they appear on similar scales.
In this work, we assume the non-linear alignment model \citep{nla1, nla2}.
There, the IA amplitude is given by
\begin{equation}
    A_{\mathrm{IA}} (z) = A_{\mathrm{IA}} \Bigg[ \frac{(1+z)}{(1+z_{\mathrm{piv}})} \Bigg]^{\eta_{\mathrm{IA}}}
\end{equation}
with a pivot redshift $z_{\mathrm{piv}}$, the amplitude $A_{\mathrm{IA}}$, and the power law slope $\eta_{\mathrm{IA}}$. 
The pivot redshift is typically chosen so that the uncertainties in $A_{\mathrm{IA}}$ and $\eta_{\mathrm{IA}}$ are as uncorrelated as possible in the posteriors. 
IA impact the angular power spectra as follows (for details see e.g. \citealt{krause2016impact,lamman2023ia}):
\begin{align}
    C^{ij}_{\mathrm{\delta_{\mathrm{g}}}\mathrm{\kappa}}\left(\ell\right) &\longrightarrow  \, C^{ij}_{\mathrm{\delta_{\mathrm{g}}}\mathrm{\kappa}}\left(\ell\right) + C^{ij}_{\mathrm{\delta_{\mathrm{g}}}\mathrm{I}}\left(\ell\right) \, , \nonumber\\
    C^{ij}_{\mathrm{\kappa}\mathrm{\kappa}}\left(\ell\right) &\longrightarrow \, C^{ij}_{\mathrm{\kappa}\mathrm{\kappa}}\left(\ell\right) + C^{ij}_{\mathrm{\kappa}\mathrm{I}}\left(\ell\right) + C^{ij}_{\mathrm{I}\mathrm{I}}\left(\ell\right) \, .
\end{align}

\section{\label{sc:flagship}Flagship simulation}

\begin{figure}
\centering
\includegraphics[angle=0,width=1.0\hsize]{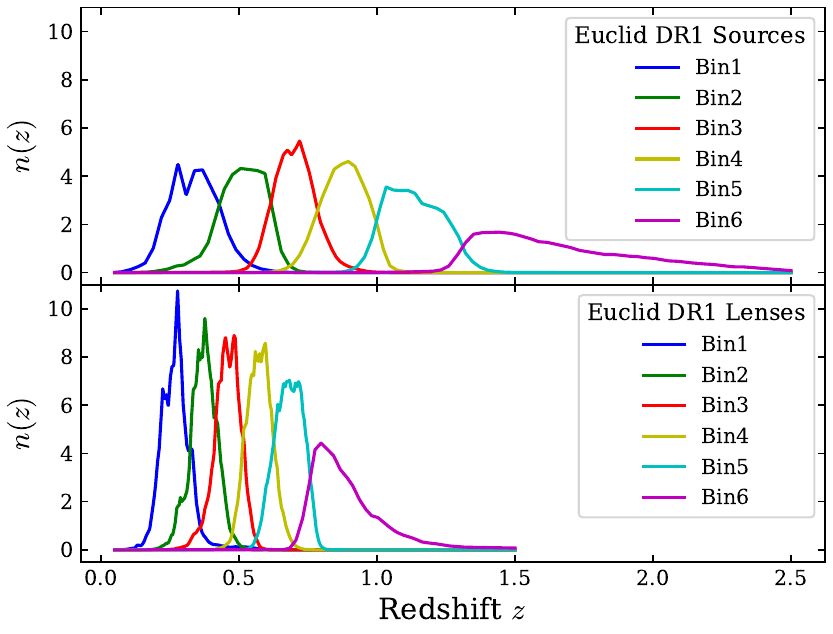}
\caption{Normalised redshift distributions of source galaxies (Top) and lens galaxies (Bottom) from the Flagship simulation assuming a DR1-like setup.}
\label{fig:dr1-nz}
\end{figure}

Along with an upcoming wealth of observational data, the Euclid Collaboration also produced a massive simulated galaxy catalogue, known as the \Euclid Flagship mock galaxy catalogue \citep{EuclidSkyFlagship} publicly available on the CosmoHub website\,\footnote{\url{https://cosmohub.pic.es/catalogs/353}} \citep{carretero2017cosmohub, tallada2020cosmohub}. This simulation closely resembles the galaxies observed by the \Euclid satellite. 
It was created using a particle-based simulation called \texttt{PKDGRAV3} \citep{Potter_2017}, which tracked the movement of billions of particles in a simulated universe.
The assumed underlying cosmology is a flat $\Lambda$CDM model.
Galaxies were assigned to specific dark matter haloes using a combination of halo occupation distribution and abundance matching techniques.
The dark matter halo identification was performed using \texttt{ROCKSTAR} \citep{behroozi2012rockstar}.
The simulation includes various galaxy properties, such as luminosity, redshift, position, velocity, shape, size, stellar mass, star formation rate, and lensing properties. 
It also contains simulated photo-$z$s for galaxies, based on \Euclid's capabilities, and covers a representative octant of the sky, offering a crucial testing ground for analysis methods applicable to the full Euclid Wide Survey.
Overall, the Flagship simulation is a valuable tool for data processing and scientific analysis, supporting the \Euclid mission and future research. 
In this work, we assume the Flagship cosmology and redshift distributions as baseline (see Table~\ref{t:lambdacdm parameters} and Fig.~\ref{fig:dr1-nz} for details).

For both the lens and source redshift distributions, we use six equipopulated bins with magnitude cuts applied at $\IE\leq$ 22.25 and $\IE\leq$ 24.5, respectively.
The redshift distributions have a minimum photo-$z$ value of $z$ = 0.2 and a maximum value of $z$ = 2.5.
Both redshift distributions are depicted in Fig.~\ref{fig:dr1-nz}.

Given that our analysis is focused on DR1, we do not include the full Flagship area but select an area of 2500\,$\deg^2$, corresponding approximately to early expectations
for the DR1 area \citep{EuclidSkyOverview}. 

Although most of the tests in this work are performed using simulated theory data vectors, we also investigated cases including Flagship measurements. These will be presented in Appendix ~\ref{app_flagship}.

\section{\label{sc:Method}Methodology}

\subsection{Parameter space and priors}
\begin{table}
    \centering
    \caption{Cosmological and astrophysical or systematic parameters, as well as priors used in the analyses. 
    }
    \vspace{5mm}
   \begin{tabularx}{0.485\textwidth}{c c c}
   \hline
    Parameter & Fiducial & Prior\\\hline
    & \textbf{Cosmology} & \\
    $\Omega_{\mathrm{m}}$ &  0.319 & [0.1, 0.9] \\
    $h$ &  0.67 & [0.55, 0.91]\\
    $\Omega_{\mathrm{b}}$ &   0.049 & [0.03, 0.07]\\
    $n_{\mathrm{s}}$ & 0.96 & [0.87, 1.07]\\
    $A_{\mathrm{s}}10^{9}$ & 2.1 & [0.5, 5.0]\\
    $w_{\mathrm{0}}$ & $-$1.0 & [$-$2.0, $-$0.333] \\
    $w_{a}$ & 0.0 &  [$-$3.0, 3.0]\\
    \hline
     & \textbf{Linear galaxy bias} & \\
     $b^1$ & 1.086 & [0.8, 3.0] \\
     $b^2$ & 1.279 & [0.8, 3.0] \\
     $b^3$ & 1.284 & [0.8, 3.0] \\
     $b^4$ & 1.400 & [0.8, 3.0] \\
     $b^5$ & 1.485 & [0.8, 3.0] \\
     $b^6$ & 1.789 & [0.8, 3.0] \\
     \hline
     & \textbf{Magnification bias} & \\
     $\alpha^1$ & 0.69 & Fixed \\
     $\alpha^2$ & 0.68 & Fixed \\
     $\alpha^3$ & 0.79 & Fixed \\
     $\alpha^4$ & 0.89 & Fixed \\
     $\alpha^5$ & 1.20 & Fixed \\
     $\alpha^6$ & 2.18 & Fixed \\
     \hline
    & \textbf{Intrinsic Alignment} & \\
    $z_{\mathrm{piv}}$ & 0.62 & Fixed \\
    $A_{\mathrm{IA}}$ & 1.0 & [$-$5.0, 5.0]\\
    $\eta_{\mathrm{IA}}$ & 0.0 & [$-$5.0, 5.0] \\
    \hline
     & \textbf{Shear calibration} & \\
     $m^i$ & 0.0 & $\mathcal{N}$(0.0, 0.005) \\
    \hline
     & \textbf{Source photo-$z$ uncertainties} & \\
     $\Delta z_{\mathrm{s}}^1$ & 0.0 & $\mathcal{N}$(0.0, 0.0027) \\
     $\Delta z_{\mathrm{s}}^2$ & 0.0 & $\mathcal{N}$(0.0, 0.0030) \\
     $\Delta z_{\mathrm{s}}^3$ & 0.0 & $\mathcal{N}$(0.0, 0.0034) \\
     $\Delta z_{\mathrm{s}}^4$ & 0.0 & $\mathcal{N}$(0.0, 0.0038) \\
     $\Delta z_{\mathrm{s}}^5$ & 0.0 & $\mathcal{N}$(0.0, 0.0043) \\
     $\Delta z_{\mathrm{s}}^6$ & 0.0 & $\mathcal{N}$(0.0, 0.0052) \\
     \hline
\end{tabularx}
\tablefoot{The fiducial parameters are taken from the Flagship simulation, whereas the priors on the cosmology, galaxy bias, IA, and magnification bias are based on recent analyses from DES \citep{abbott2022dark,2023desextensions}. The flat and Gaussian priors on the multiplicative shear bias and $n(z)$ uncertainties are defined by the Euclid Collaboration \citep{EuclidSkyOverview}.}
    \label{t:lambdacdm parameters}
\end{table}
We investigate the impact of uncertainties in the lens redshift distribution considering a flat universe with cold dark matter and dark energy.
Tests are conducted on two cosmological models, namely $\Lambda$CDM and $w_{\mathrm{0}}w_{a}$CDM.
In both cases, the fiducial values for the cosmological parameters considered are taken directly from the Flagship simulation. 
The parameters describing the background cosmology are 
\Omm, and the dimensionless Hubble constant $h$ with $H_{\rm 0}$\,=\,100\,$h$\,km\,s$^{-1}$\,Mpc$^{-1}$.
We also require parameters that describe the regime of linear perturbation, that is, the baryon density parameter \Omb, the slope of the primordial power spectrum $n_{\rm s}$, and one parameter related to the amplitude of matter fluctuations, $A_{\rm s}$.
In $w_{\mathrm{0}}w_{a}$CDM this set of parameters is extended to the dark energy equation-of-state parameters $w_{\mathrm{0}}$ and $w_{a}$.
The cosmological model used here includes the dark energy equation-of-state parameters following the Chevallier--Polarski--Linder parametrisation \citep{CHEVALLIER_2001, Linder_2003}
\begin{equation}
    w(z)\,=\,w_{\rm 0} + w_{a} \frac{z}{1 + z} \; .
\end{equation}

In addition, we include a number of astrophysical and systematic parameters relating to the galaxy bias, IA, and shear bias described in Sect. \ref{subs:systematics}.
An overview of the cosmological, astrophysical, and systematic parameters used and the priors on them is given in Table \ref{t:lambdacdm parameters}.
We note that the setup for forecasts used in \citet{EuclidSkyOverview} differs from the one used in our work as we focus specifically on DR1.
The priors on the cosmology, the linear galaxy bias, and IA are assumed based on recent analyses from DES, and are adopted from \cite{abbott2022dark,2023desextensions}. 
Moreover, we assume priors for the systematic effects of the source redshift distribution and shear calibration defined by the Euclid Collaboration \citep{EuclidSkyOverview}.
We note that the IA priors used in \citet{EuclidSkyOverview} are narrower than the ones we used. We tested the impact of the wide IA priors from DES and found that the reduction of the prior width to the values used in \citet{EuclidSkyOverview} does not affect our results significantly. For instance, the outcome of the median value of the parameter $w_{\rm 0}$ was changed by 0.003\,$\sigma$, the median of $w_{a}$ by 0.002\,$\sigma$. This confirms that our choice of IA priors does not impact the key conclusions of this analysis.

In this work, we investigate uncertainties on the mean and width of the lens $n(z)$. 
The exact parametrisations of the uncertainties are given in Eq. \eqref{meanuncertainty} and Eq. \eqref{widthuncertainty}, respectively. 
The mean redshifts in the six lens redshift bins are \{0.273, 0.372, 0.458, 0.574, 0.681, 0.816\}.

For the forecast we focus on the impact of photo-$z$ uncertainties on the cosmological parameters $\Omega_{\mathrm{m}}$, $\sigma_{\mathrm{8}}$, and the related parameter $S_{\mathrm{8}}$, 
\begin{equation}
    S_{\mathrm{8}} = \sigma_{\mathrm{8}} \sqrt{\frac{\Omega_{\mathrm{m}}}{0.3}} \; ,
\end{equation}
since those are the most constrained parameters in a $3\times2$pt analysis. 
We also evaluate the results of the chains based on their constraining power on the dark energy parameters $w_{\mathrm{0}}$ and $w_{a}$.

\subsection{Covariances}\label{subs covs}

The covariance matrix is an essential quantity relevant for determining the uncertainties of the cosmological parameters.
The main analysis used in this work is carried out using an analy\-tical covariance.
We generate them 
for our \Euclid DR1 setup using \texttt{CosmoCov} \citep{CosmoCov}, a code package based on the \texttt{CosmoLike} framework \citep{CosmoLike}.
In this work, we use only the Gaussian matrices and do not include the non-Gaussian covariance because we focus on the forecasting and comparison of the constraining power of different scenarios.
The number densities used for the computation of the covariance are $N_{\mathrm{gal, l}}^i = 0.558$ and $N_{\mathrm{gal, s}}^i = 4.05$ galaxies/arcmin$^2$ for each tomographic bin of the lens and source samples, respectively.

\subsection{Scale cuts}\label{subs scalecuts}
The aforementioned modelling of systematic effects is only valid on certain scales.
Effects of non-linear and stochastic galaxy bias or baryonic physics on the matter power spectrum were not accounted for in the analysis, but instead their impact has been greatly diminished by applying scale cuts.

Baryonic feedback mostly affects small-scale cosmic shear mo\-delling whereas non-linear bias is known to be a major contamination in galaxy clustering and galaxy-galaxy lensing analyses.
We mitigate the contamination caused by these effects by removing the data points that are significantly affected \citep{scalecuts1, scalecuts2}.
Given that we focus on analysing noiseless theory data vectors generated with our baseline model, scale cuts are not needed in order to obtain unbiased cosmological constraints. However, to have realistic forecasts, we assume scale cuts based on the DES-Y3 3$\times$2pt analysis \citep{krause2021dark,abbott2022dark}. In particular, we use their comoving scale cuts for the non-linear galaxy bias model, $R$ = 4\,Mpc\,$h^{-1}$, for galaxy clustering and galaxy-galaxy lensing, 
assuming that the modelling based on \Euclid DR1 data will be able to probe the same scales as the precursor analysis from DES.
The angular scale cuts for the $i$-th bin are then calculated as 
\begin{equation}
    \theta_{\mathrm{min}}^{i}\,=\,\frac{R}{\chi(\,z_{\mathrm{mean}}^{i})} \, ,
\label{eq:scale_cut}
\end{equation}
with the mean redshift in each lens bin $z_{\mathrm{mean}}^{i}$ 
as well as the minimum comoving transverse scale $R$.
The scale cuts for cosmic shear are also based on the DES-Y3 analysis and were extended to the six bins present in our source sample by using the same values as the fourth and thus highest redshift bin of DES-Y3. 

We note that in Appendix \ref{app_flagship} we use a different set of scale cuts for $w(\theta)$ in order to be able to fit the Flagship measurements with a linear galaxy bias model.

\subsection{Likelihood exploration and forecasting}\label{subsec:likefish}

We perform our analysis in the context 
of a flat $\Lambda$CDM framework as well as a $w_{\mathrm{0}}w_{a}$CDM cosmology.
Most of our analysis consists of likelihood explorations using Monte Carlo methods with simulated theory data vectors.
These theory data vectors are theoretical predictions evaluated at the fiducial cosmology.
Both the theory data vectors and the Monte Carlo samples are generated using the \texttt{CosmoSIS}\footnote{\url{https://cosmosis.readthedocs.io/}} framework \citep{CosmoSIS}.
In parti\-cular, for the sampling we use the publicly available \texttt{nautilus} sampler \citep{Nautilus}, a machine learning-
enhanced sampler utili\-sing importance nested sampling to compute the Bayesian posteriors and estimate the evidence. We refer to \citet{lewis2000efficient}, \citet{howlett2012cmb}, and \citet{mcewen2016fast} for the postprocessing of the chains.

Bayesian parameter inference involves several key components: a data set $\hat{\Vec{D}}$, in this work represented by the simulated theory data vector, a theoretical model $\Vec{T_{\mathrm{M}}}(\Vec{p})$ with parameters $\Vec{p}$, the priors on the model $M$, as well as the covariance $\tens{\Gamma}$.
\noindent The theory data vector has the form 
\begin{equation}
    \hat{\Vec{D}}\,=\,\left\{\hat{w}^{i}(\theta),\,\hat{\gamma}_{\mathrm{t}}^{ij}(\theta),\,\hat{\xi}_{\mathrm{+/-}}^{ij}(\theta)\right\} 
\end{equation}
for the full 3$\times$2pt analysis and is compared to a theoretical model prediction
\begin{equation} \label{eq:estimvec}
    \Vec{T_{\mathrm{M}}}(\Vec{p})\,=\,\left\{w^{i}(\theta, \Vec{p}),\,\gamma_{\mathrm{t}}^{ij}(\theta, \Vec{p}),\,\xi_{\mathrm{+/-}}^{ij}(\theta, \Vec{p})\right\} \; .
\end{equation}
We assume a Gaussian likelihood that includes both theory vector and theory prediction and can be expressed as
\begin{equation}
  \mathcal{L}(\hat{\Vec{D}}|\Vec{p}, M) \propto \text{exp}\left\{-\frac{1}{2} \left[ (\hat{\Vec{D}} - \Vec{T_{\mathrm{M}}}(\Vec{p}))^\mathrm{T} \tens{\Gamma}^{-1} (\hat{\Vec{D}}- \Vec{T_{\mathrm{M}}}(\Vec{p})) \right] \right\} \; .
\end{equation}

The posterior probability distributions for the individual parameters are inferred via Bayes' theorem
\begin{equation}
    P(\Vec{p}|\hat{\Vec{D}}, M) \propto \mathcal{L}(\hat{\Vec{D}}|\Vec{p}, M) \ \Pi(\Vec{p}|M)\,,
\end{equation}
with the prior probability distribution on the parameters $\Pi(\Vec{p}|M)$.

In order to study the impact of lens photo-$z$ uncertainties on the cosmological constraints, we quantify the impact on the constraints through shifts to the mean of the posteriors and changes in constraining power.
The constraining power of a given pair of parameters is estimated by using the 2D figure of merit (FoM, \citealt{force2006report, wang2008figure}) defined as 
\begin{equation}
    \mathrm{FoM}_{p_{\rm 1}p_{\rm 2}} = \sqrt{\text{det\,} \Big(\tens{C}^{-1}_{p_{\rm 1}p_{\rm 2}}\Big)}
\end{equation}
for two parameters $p_{\rm 1}$ and $p_{\rm 2}$ with $\tens{C}^{-1}_{p_{\rm 1}p_{\rm 2}}$ being the inverse of the covariance matrix of these parameters.
As per definition, the FoM is proportional to the inverse of the area of the confidence ellipse in the plane of the parameters $p_{\rm 1}$ and $p_{\rm 2}$.
Therefore, a decrease in FoM indicates a loss of constraining power. 

A computationally faster way of forecasting the expected constraining power is rooted in the Fisher matrix formalism. Although it assumes that both the posterior and the likelihood can be characterised by a multivariate Gaussian function at their maximum, it can be considered a good approximation in several cases. The Fisher matrix $F_{p_1p_2}$ can be computed in terms of the derivatives of the angular correlations defined above and their covariance, as described in \citet{2009arXiv0906.0664H},
\begin{equation}
F_{p_1p_2}= \left(\frac{\partial \Vec{T_{\mathrm{M}}}}{\partial p_1}\right)^T\ \tens{\Gamma}^{-1}\ \left(\frac{\partial \Vec{T_{\mathrm{M}}}}{\partial p_2}\right)\,, 
\end{equation}
where the covariance of the estimators, $\Gamma^{km}(\theta,\theta')={\rm Cov}\left[{T_{\mathrm{M}}^k}(\theta),{T_{\mathrm{M}}^m}(\theta')\right]$, is determined at the fiducial value of the parameters $p_i$ and ranges through all the estimators and angular bins in the data vector (Eq.~\ref{eq:estimvec}). In this work, we also use \texttt{CosmoSIS} \citep{CosmoSIS} to compute the Fisher matrices. In practice, the Fisher matrix $F_{p_1p_2}$ is a Gaussian approximation of the shape of the posterior close to the the maximum. If we assume Gaussian priors, they can be expressed as an additive term to the Fisher matrix.
The Fisher matrix formalism can also be utilised to estimate the bias on the best fit if the wrong model is applied. While this estimation may come more easily in a Monte Carlo analysis, in the Fisher formalism we have to assume nested models, that is, a model with parameters $\Vec{p}$ can be treated as part of a larger model $\Vec{q}$ such that $\Vec{p}\in \Vec{q}$. For the purpose of this paper, this assumption suffices. A Fisher matrix $H_{q_1q_2}$ can be defined for the set of parameters $\Vec{q}$. Note that if there are $n$ parameters $p_i$ and $m$ parameters $q_j$, then $m>n$ and the first $n$ parameters of $\Vec{q}$ are selected to be the same as $\Vec{p}$. Similarly, $\tens{F}$ is a subset of $\tens{H}$. Then, the bias on the best fit of the subset model $p_i$ by fixing a larger model with the wrong parameters $q_{n+1},\ldots,q_m$ is given by \citep{2021MNRAS.504..267F}
\begin{equation}
    \delta p_i=- \delta q_k~ {H}_{q_k p_j}\ {F}^{-1}_{p_j p_i}\,.
\end{equation}

In this work, we rely on Monte Carlo methods in Sects.~\ref{mc} and Appendix \ref{app_flagship}, but also use the Fisher matrix approach in Sect.~\ref{sbc:explo_fisher} to explore a wider variety of scenarios.

\section{\label{sc:Results}Results}
In this section we present the main findings of our analysis regarding the impact of uncertainties in the lens redshift distribution on the \Euclid DR1 photometric galaxy clustering and 3$\times$2pt analy\-ses.
We assume -- in most cases -- perfect knowledge of the source redshift distributions. 
The main part of this work consists of Monte Carlo analyses of several scenarios.
In particular, we present cases in which we marginalise over the photo-$z$ uncertainties in Sect. \ref{mc} and compare these against Fisher forecasts in Sect. \ref{sbc:explo_fisher}.
To further test the individual impacts of shift and stretch, Sect.~\ref{biasess} describes cases in which we directly shift or stretch the lens redshift distribution when analysing the simulated data (assuming no uncertainty in the redshift distributions).

\subsection{\label{mc}Loss of constraining power}

In the following we compare the cosmological constraints obtained from the analyses of simulated theory data vectors with different $n(z)$ assumptions performing a full Monte Carlo simulation of the 
3$\times$2pt analysis. 
In this section, we analyse the simulated data vector with the true lens redshift distribution, but margina\-lising over uncertainties on the mean ($\Delta z$) and width ($\sigma_z$).
We then quantify the loss of constraining power via the FoM values.

\subsubsection{\label{sbc:loss_fom}Exploration with Monte Carlo sampling}

For this part of the analysis, we considered a baseline $w_{\mathrm{0}}w_{a}$CDM data vector that includes the fiducial values for IA and the dark energy equation-of-state parameters from Table~\ref{t:lambdacdm parameters}.
\Euclid requirements for the source redshift systematic effects are based on studies in \citet{kitching2008systematic}.
In their ana\-lysis, the authors explored the impact of uncertainties in the mean of the source redshift distributions on the constraints on $w_0$-$w_{a}$ using Gaussian priors and marginalising over all nuisance parameters.
The \Euclid requirements defined for weak lensing were based on a recovery of 80$\%$ of the $w_{\mathrm{0}}w_{a}$ FoM.
In contrast to the Fisher analysis used in \mbox{\citet{kitching2008systematic}}, here we use Monte Carlo methods and consider a \Euclid DR1 3$\times$2pt analysis, supported by Fisher forecasts for certain cases.

\begin{figure*}
\centering
\includegraphics[angle=0,width=0.92\hsize]{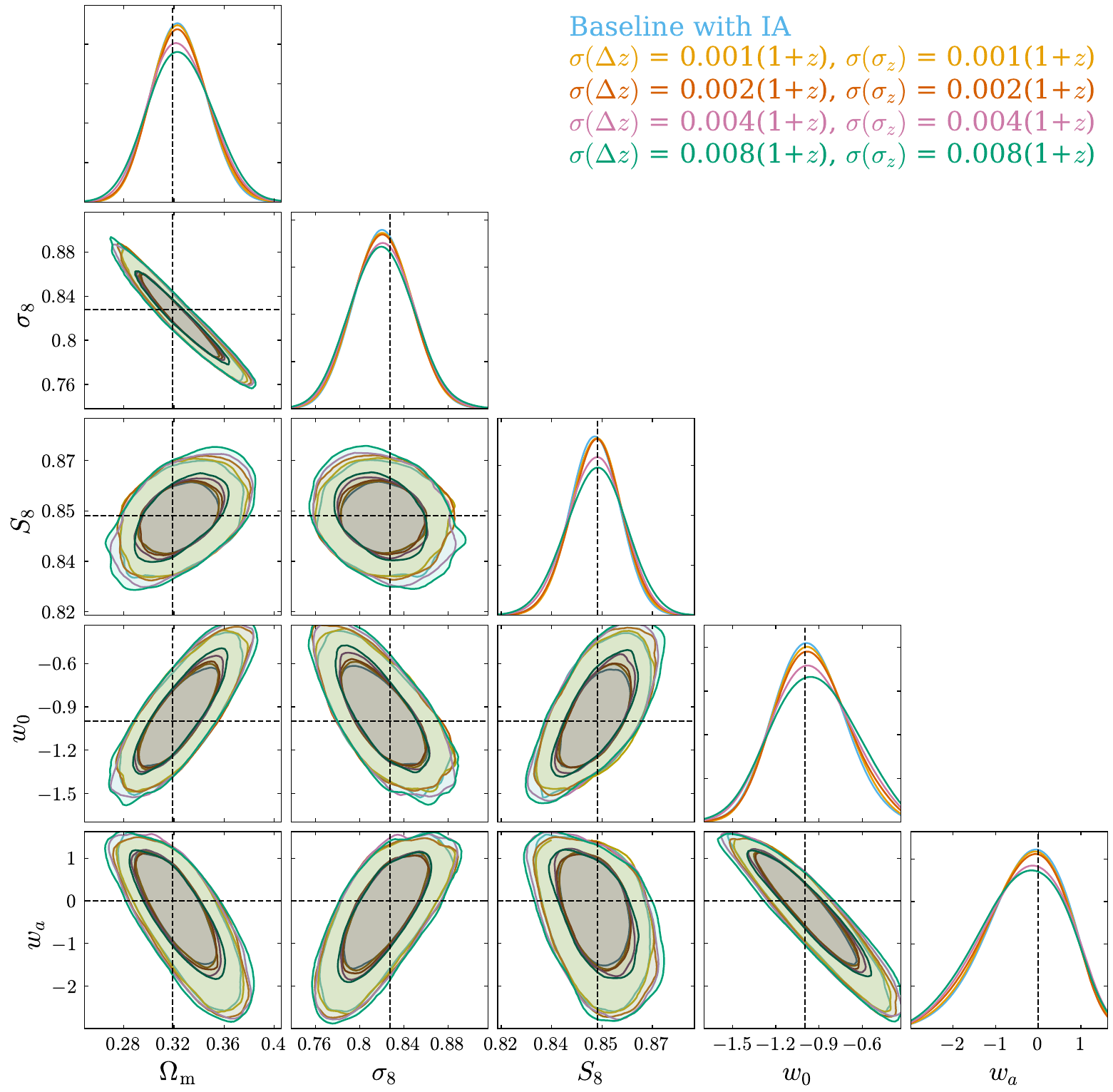}
\caption{Cosmological constraints on selected cosmological parameters obtained from the 3$\times$2pt analysis of a simulated theory data vector with the true redshift distribution, marginalising over uncertainties on the mean ($\Delta z$) and width ($\sigma_z$) of the lens redshift distributions and assuming different standard deviations $\sigma$ of the corresponding Gaussian priors. The \lq baseline\rq case corresponds to the no-marginalisation of lens photo-$z$ parameters. As the width of the Gaussian priors increases, the $w_{\mathrm{0}}w_{a}$ FoM degrades from $\sim$5$\%$ (orange contours) up to $\sim$30$\%$  (green). There is a 11$\%$ degradation for the default requirement for the sources combined with the corresponding stretch (red). 80$\%$ of the FoM are recovered for the pink contours, corresponding to prior sizes of $\sigma$ = 0.004(1+$z$) on both mean and width.}
\label{fig:gaussian_contours}
\end{figure*}

The FoM percentages with respect to the baseline for three different parameter combinations are listed in Table \ref{t:FoMgaussian} and the corresponding posterior distributions are shown in Fig.~\ref{fig:gaussian_contours}. 
Overall, the FoMs of all parameter combinations degrade by similar percentages.
We recover approximately 80$\%$ of the $w_{\mathrm{0}}w_{a}$ FoM in the case of the pink contours, which assume a 0.4$\%$ uncertainty in the mean and 
in the width of the lens redshift distribution.
Since \Euclid aims to constrain $w_{\mathrm{0}}$ at the 2$\%$ level and $w_{a}$ at the 10$\%$ level \citep{EuclidSkyOverview}, any further degradation of the FoM would considerably reduce the possibility of \Euclid to shed light on the time evolution of dark energy.

Figure \ref{fig:foms} illustrates how the FoM changes for the parameter combinations $w_{\mathrm{0}} \times w_{a}$ (green curve) and $\Omega_{\rm m}\times S_8$ (shown in blue), clearly showing a reduction in constraining power compared to the baseline scenario.  
The similar trend across both cases indicates that this loss is correlated with the uncertainty in the lens redshift distribution parameters. 
As early work in this field highlighted, the requirement on the mean inherently constrains the knowledge of the distribution's width \citep{ma2008size}. 
When calibrating redshift distributions with a finite spectroscopic sample, the requirements on the mean inherently constrain the knowledge of the distribution’s width. In such calibration scenarios, the uncertainties in the mean and width scale together. It is important to note, however, that this strong correlation depends on the specific source of the $n(z)$ uncertainties. Other systematic effects, such as mis-estimation of photometric noise or unmodelled Galactic extinction, could theoretically affect the width independently of the mean. Nevertheless, for the calibration strategies and the peaked distributions expected in a realistic \Euclid DR1 scenario considered here, we expect the mean and width uncertainties to be clearly interdependent.

\subsubsection{\label{sbc:explo_fisher}Exploration with Fisher matrices}

So far we have used synthetic data vectors to assess how cosmological constraints are sensitive to systematic uncertainties. However, while full Monte Carlo analyses are the proper way to map the shape of the posterior with data, they are also time consuming and require significant computational effort. A faster approach, in lack of observational or simulated data, is to use the Fisher matrix formalism to approximate the covariance of the parameters assuming Gaussian posteriors and likelihoods.

As described in Sect.~\ref{subsec:likefish}, we compute the Fisher matrices using \texttt{CosmoSIS} with the fiducial values for the parameters defined in Table~\ref{t:lambdacdm parameters}. We chose the {\it smooth} method using the default settings of \texttt{CosmoSIS} to compute the numerical derivatives. We varied the settings to confirm that our choice was appropriate and the forecasts were stable. 
Furthermore, we follow \cite{2017MNRAS.464.4747C} and \cite{2021PhRvD.103d3503P} and, for each parameter with flat priors in Table~\ref{t:lambdacdm parameters}, we add an additional Gaussian prior centred at the fiducial value and with standard deviation $\sigma_p=(p_{\rm max}-p_{\rm min})/2$ (with $p_{\rm max}$ and $p_{\rm min}$ being the extremes of the parameter ranges in Table~\ref{t:lambdacdm parameters}). This is justified because the Fisher matrix fails to estimate the posterior distributions in the presence of non-Gaussian posteriors. The previous authors have found it to lead to a good agreement between a Monte Carlo analysis and the Fisher formalism. For those parameters with Gaussian priors in the Monte Carlo analysis, we add a Gaussian

\begin{table*}
    \centering
    \caption{FoM percentages for several parameter combinations.}
    \vspace{5mm}
   \begin{tabularx}{0.53\textwidth}{c c c  }
   \hline
    Chain & $w_{\mathrm{0}}\times w_{a}$ & 
    $\Omega_{\mathrm{m}}\times S_{\mathrm{8}}$ 
    \\\hline
    $\sigma (\Delta z)$ = 0.001(1+$z$), $\sigma (\sigma_z)$ = 0.001(1+$z$) &  
    96 & 
    95 
    \\
    $\sigma (\Delta z)$ = 0.002(1+$z$), $\sigma (\sigma_z)$ = 0.002(1+$z$) &  
    89 &  
    91 
    \\
    $\sigma (\Delta z)$ = 0.004(1+$z$), $\sigma (\sigma_z)$ = 0.004(1+$z$) & 
    80 & 
    83  
    \\
    $\sigma (\Delta z)$ = 0.008(1+$z$), $\sigma (\sigma_z)$ = 0.008(1+$z$) & 
    73 & 
    76  
    \\
    \hline
    \end{tabularx}
    \tablefoot{The FoMs were obtained in the 3$\times$2pt analysis of a simulated theory data vector with the true redshift distribution, marginalising over uncertainties on the mean ($\Delta z$) and width ($\sigma_z$) of the lens redshift distributions and assuming different standard deviations $\sigma$ of the corresponding Gaussian priors. 
    The numbers quoted are the percentage of the FoM recovered  with respect to the \lq baseline\rq \,case that corresponds to the non-marginalisation of lens photo-$z$ parameters.}
    \label{t:FoMgaussian}
\end{table*}

\clearpage
\begin{figure}
\centering
\includegraphics[angle=0,width=1.0\hsize]{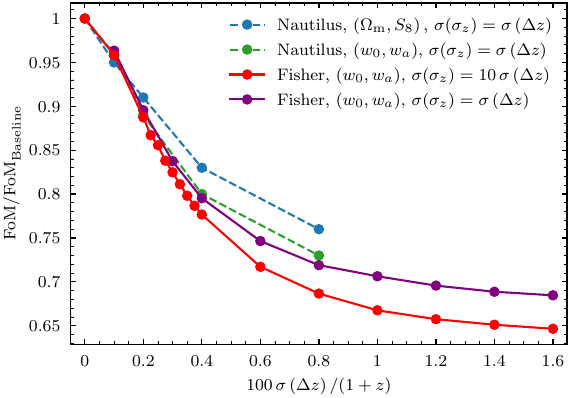}
\caption{Evolution of the FoM of two selected parameter combinations considering different priors on the lens redshift distribution parameters with respect to the baseline without prior, at $\sigma\left(\Delta z\right)=0$. The corresponding values of the FoMs from the Monte Carlo analysis are listed in Table \ref{t:FoMgaussian}. The green and purple curves correspond to the same scenario tested with the Monte Carlo (green) and Fisher (purple) approaches.} 
\label{fig:foms}
\end{figure}

\noindent  prior in the same way but assuming the same standard deviation. While a full comparison between the two approaches is given in Appendix \ref{app:comp_fish}, we note that the expected marginal errors from the Fisher matrix broadly agree with the Monte Carlo results. 
The exception is when estimating the bias on the best-fit parameters. Here, the estimates from the Fisher analysis deviate from the Monte Carlo analysis as the Gaussian approximation for the maximum of the posterior breaks down. 

Finally, in Fig. \ref{fig:foms}, we take advantage of the Fisher approach to explore more cases than those explored in Sect.~\ref{sbc:loss_fom} with the Monte Carlo analysis.
We 
estimate the loss of constraining power when marginalising over the associated nuisance para\-meters, with the same setup as in Sect.~\ref{sbc:loss_fom}. We find a similar trend in the evolution of the FoM of dark energy when varying the width of the Gaussian priors used for $\Delta z$ and $\sigma_z$. 
We find that the ratio of the degrading $w_0$$w_a$ FoM over the baseline FoM is identical within 1\% for $\sigma\left(\Delta z\right) \leq 0.002(1+z)$ in the two cases $\sigma\left(\sigma_z\right)=10\,\sigma\left(\Delta z\right)$ and $\sigma\left(\sigma_z\right)=\sigma\left(\Delta z\right)$. When $\sigma\left(\Delta z\right) \geq 0.006(1+z)$, the FoM is increased by 5\% in the realistic $\sigma\left(\sigma_z\right)=\sigma\left(\Delta z\right)$ case with res\-pect to the other scenario. 
The FoM ratio reaches 80\% for $\sigma\left(\Delta z\right)=\sigma\left(\sigma_z\right)=0.004(1+z)$. 
As anticipated from the correlated uncertainties of our calibration approach, we find that the final constraints from the 3$\times$2pt analysis are driven most exclusively by the mean. For the calibration strategies considered in this work, it is therefore sufficient to place a quality cut on the calibration of the mean, as the requisite precision on the width is implicitly satisfied given the coupling between the two parameters. 
Given that this scenario best reflects the expected physical properties of the redshift distributions, we therefore recommend reaching this level of photo-$z$ accuracy to avoid reducing significantly the constraining power of $w_0$-$w_{a}$ in the \Euclid DR1 3$\times$2pt data analysis.

\subsection{\label{biasess}Bias on inferred cosmological parameters}
Our second approach to study the impact of lens photo-$z$ uncertainties on the inferred cosmology is to analyse the simulated data with perturbed lens redshift distributions (with no marginalisation of lens photo-$z$ nuisance parameters). Since we are assuming perfect knowledge of the true redshift distribution, but analysing the data with a modification of it, the inferred posterior distributions will be biased by some amount. We also ana\-lyse the simulated data with the true redshift distribution (the so-called \lq baseline\rq) to use as a reference.
We then quantify this bias via the percentage differences of the mean values of the 1D marginalised posterior distributions of the parameters, by comparing the baseline with a case analysed with a \lq variation\rq of the $n(z)$. For a given parameter $x$, we have
\begin{equation}
    \Delta x\,=\,\frac{\overline{x}_{\mathrm{Baseline}}-\overline{x}_{\mathrm{Variation}}}{\sigma_{\mathrm{Baseline}}}\;,
\end{equation}
with the mean of the respective parameter $\overline{x}$ and the standard deviation of the parameter given by the baseline $\sigma_{\mathrm{Baseline}}$.

\begin{figure*}
     \centering
     \includegraphics[width=0.495\textwidth]{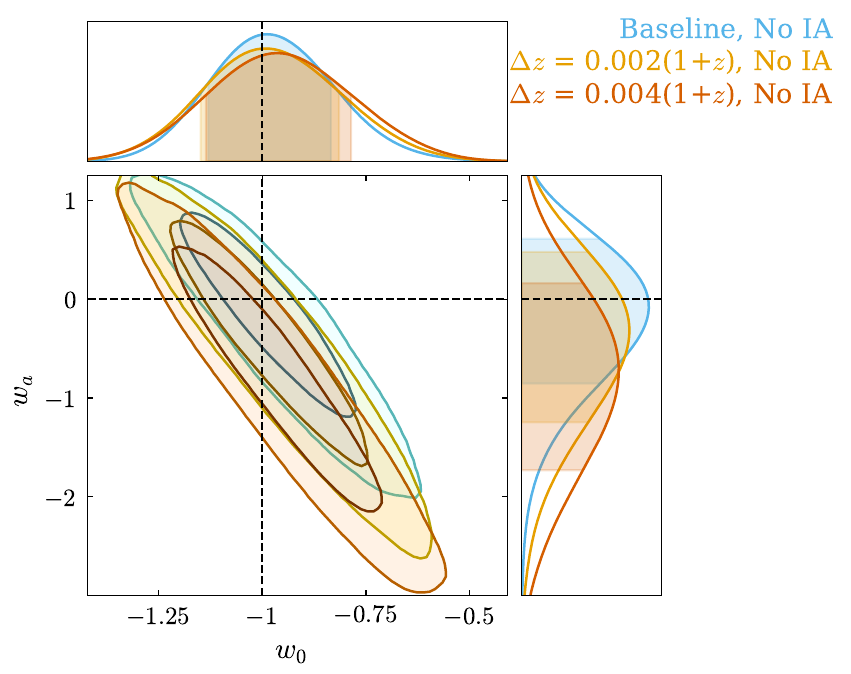}
     \hfill
     \includegraphics[width=0.495\textwidth]{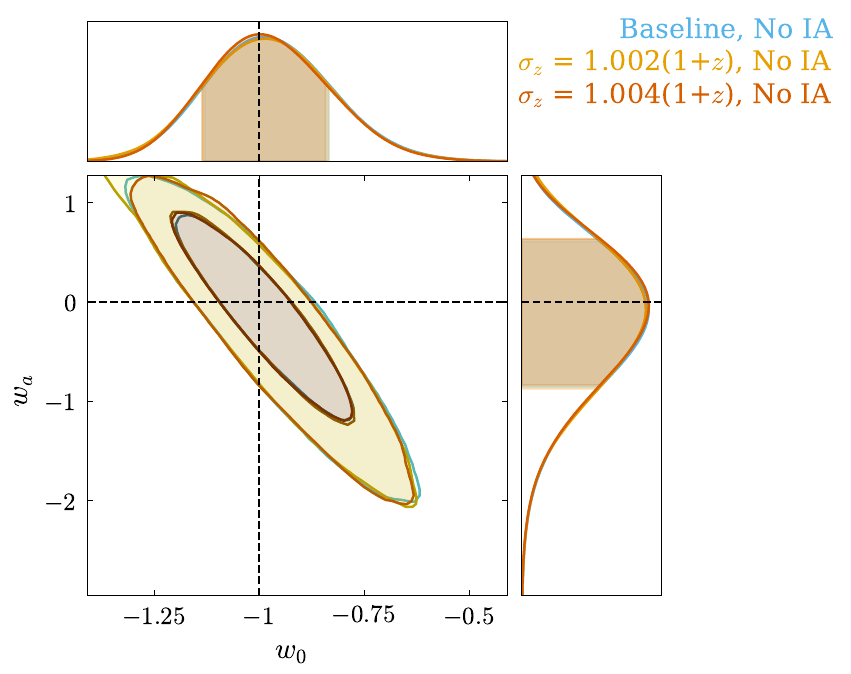}
     \caption{Comparison of cosmological constraints on $w_{\rm 0}$ and $w_{a}$ in $w_{\rm 0}w_{a}$CDM considering a 3$\times$2pt analysis. The constraints are obtained from a simulated data vector and analysed assuming different modifications of the lens redshift distribution. \textit{Left:} Comparison of the constraints from the baseline and the ones obtained from a shifted lens $n(z)$ distribution. \textit{Right:} Comparison of the constraints from the baseline and the ones obtained from a stretched lens $n(z)$ distribution.}
     \label{w0wa constraints}
\end{figure*}

\begin{figure*}
     \centering
     \includegraphics[width=0.48\textwidth]{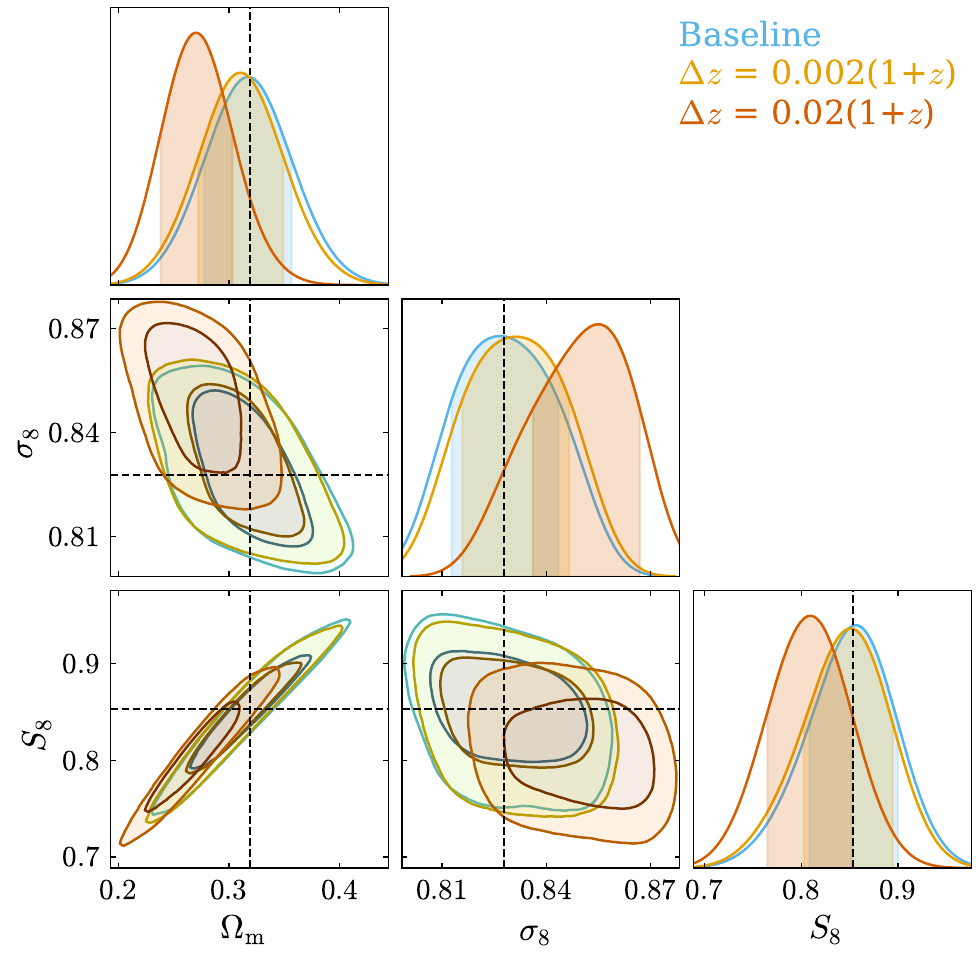}
     \hfill
     \includegraphics[width=0.48\textwidth]{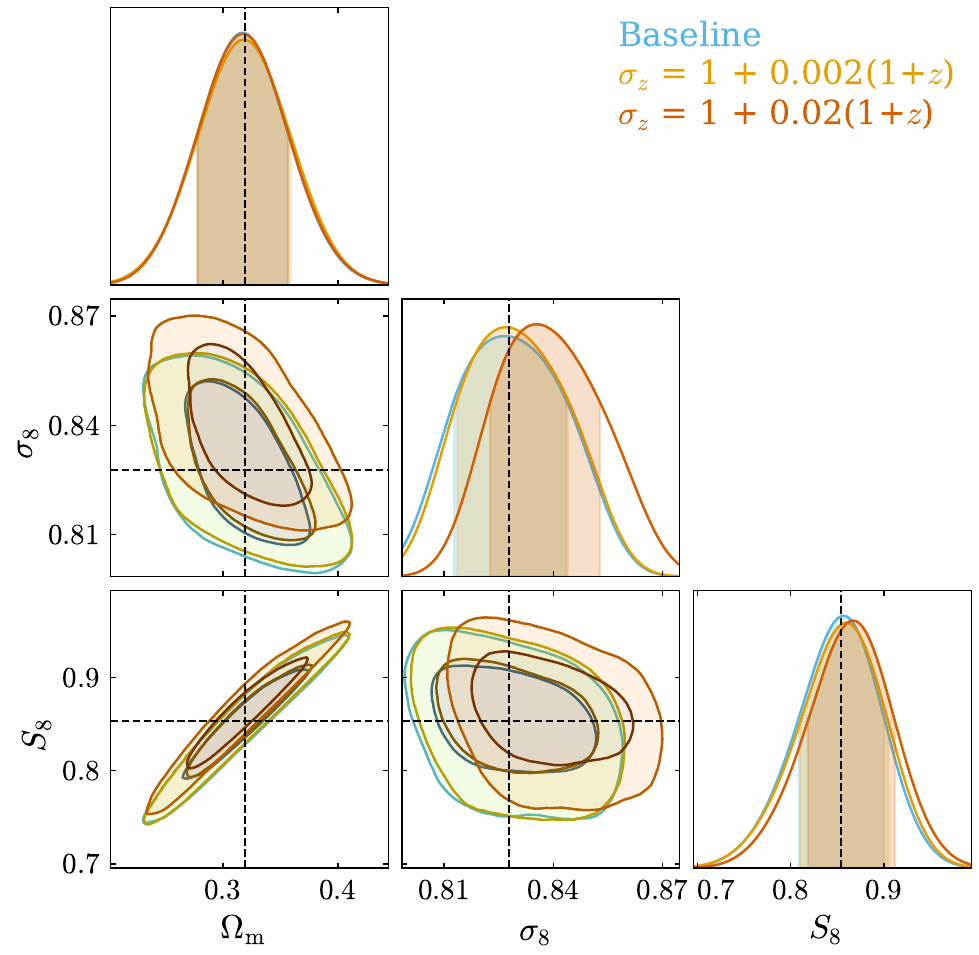}
     \caption{Comparison of cosmological constraints on $\Omega_{\mathrm{m}}$, $\sigma_{\mathrm{8}}$, and $S_{\mathrm{8}}$ in flat $\Lambda$CDM considering photometric galaxy clustering only with fixed galaxy bias. The constraints are obtained from likelihood analyses of simulated data vectors for different modifications of the lens redshift distribution. Here, and in the following contour plots, the two sets of contours indicate the 68$\%$ and 95$\%$ confidence levels, while the black dashed lines indicate the input fiducial cosmology from Table~\ref{t:lambdacdm parameters}. \textit{Left:} Comparison of the constraints from the baseline and the ones obtained from a shifted lens $n(z)$ distribution. \textit{Right:} Comparison of the constraints from the baseline and the ones obtained from a stretched lens $n(z)$ distribution.}
     \label{gcph constraints}
\end{figure*}

In order to conclude whether a given bias is significant, we follow modelling validation criteria from the DES 3$\times$2pt analyses as reference for threshold values. 
For instance, \citet{krause2021dark} assume a maximum bias of 0.3\,$\sigma$ in 2D as a threshold, while \citet{krause2017dark} use 0.5\,$\sigma$ in the 1D constraints. 
Moreover, \citet{giannini2024dark} investigated differences in the DES-Y3 2$\times$2pt cosmological constraints arising from two different lens $n(z)$ estimates, finding a shift of 0.4\,$\sigma$ in the $\Omega_{\mathrm{m}}$--\,$S_{\mathrm{8}}$ plane. 
In this work, we consider 1D biases larger than 0.5\,$\sigma$ worrisome, whereas we discuss the significance of smaller biases depending on the scenario considered. 
The impact of using modified redshift distributions on the constraints is investigated for photometric galaxy clustering only, as well as for the 
3$\times$2pt analysis. 
In some cases, the assumed stretches $\sigma_z$ are larger than the shifts to the mean by an order of magnitude, based on the trend observed in the fiducial priors for the DES-Y3 analysis \citep{Porredon2022}. 
We show the results in the form of contour plots with the 68$\%$ and 95$\%$ confidence regions for selected cases in $w_{\rm 0}w_{a}$CDM and $\Lambda$CDM in Figs. \ref{w0wa constraints} and \ref{gcph constraints}. 
When comparing these scenarios, it is important to note that the isolated 0.2$\%$ shift in the mean and the 2$\%$ stretch in the width represent different regimes of uncertainty and calibration precision. While the 0.2$\%$ shift represents a strict quality requirement, the 2$\%$ stretch illustrates a weaker constraint with significantly larger uncertainties.

\paragraph{3$\times$2pt $w_{\rm 0}w_{a}$CDM:}First, we consider a similar scenario as in Sect. \ref{mc} and examine the impact of uncertainties of the lens redshift distribution within a $w_{\rm 0}w_{a}$ dark energy model. 
The presence of IA and source systematic effects was neglected in these tests. 
For almost all cases the biases exceed 0.5\,$\sigma$ for at least one parameter considered, a level that could compromise \textit{Euclid}'s science objectives. 
In summary, none of the photo-$z$ biases considered with the $w_{\rm 0}w_{a}$CDM model meets our accuracy requirements (biases lower than 0.5\,$\sigma$) for a robust DR1 analysis, as can be inferred by Fig.~\ref{w0wa constraints} and the values given in Table \ref{t:3x2pt w0wa sigma errors}. 
As is evident from the contours in Fig. \ref{w0wa constraints}, the width of the distribution is less relevant, and the effect of shifts of the mean dominates the biases in the cosmological constraints. 
However, when the dark energy equation-of-state parameters are con\-sidered, all modifications to the lens redshift distribution lead to significant biases in the estimation of both $w_{\mathrm{0}}$ and $w_{a}$. 
When the stretches are considered, then it is important to note that those of a similar order of magnitude as the shifts do not impact the parameters significantly. However, when the stretches are increased by an order of magnitude, the accuracy requirements are not met anymore. 
This highlights the importance of accurately determining both the mean and the shape of the lens $n(z)$.
In the upcoming real-data analysis, those uncertainty parameters need to be marginalised over, which will lead to a loss of constraining power, as already discussed in Sect. \ref{mc}. 
\begin{table}
    \centering
    \caption{Biases on the 1D marginalised constraints for the 3$\times$2pt tests in a $w_{\rm 0}w_{a}$CDM model.}
    \vspace{5mm}
   \begin{tabularx}{0.495\textwidth}{c c c c c c}
   \hline
    Chain & $\Omega_{\mathrm{m}}$[$\sigma$] & $\sigma_{\mathrm{8}}$[$\sigma$] & $S_{\mathrm{8}}$[$\sigma$] &  $w_{\mathrm{0}}$[$\sigma$] &  $w_{a}$[$\sigma$]\\ \hline
    shift 0.002(1+$z$) & 0.54 & 0.32 & 0.57 & 0.004 & 0.39 \\
    shift 0.004(1+$z$) & 1.02 & 0.59 & 1.14 & 0.13 & 0.94 \\
    stretch 1.002(1+$z$) & 0.0002 & 0.002 & 0.008 & 0.04 & 0.03 \\
    stretch 1.004(1+$z$) & 0.02 & 0.02 & 0.03 & 0.03 & 0.03 \\
    stretch 1.02(1+$z$) & 0.30 & 0.28 & 0.06 & 0.59 & 0.49 \\
    stretch 1.04(1+$z$) & 0.37 & 0.39 & 0.23 & 0.89 & 0.75 \\
    \hline
\end{tabularx}
\tablefoot{Here, and in all following tables, the biases are expressed in units of the standard deviation values of the posterior distribution of the baseline chain.}
    \label{t:3x2pt w0wa sigma errors}
\end{table}

\paragraph{3$\times$2pt $\Lambda$CDM:}
In a next step, we investigate similar cases for a $\Lambda$CDM scenario.
Overall, the results we obtain are similar to those observed in the 3$\times$2pt scenarios above, with the shifts in the distribution having a greater impact on $\Omega_{\mathrm{m}}$, $\sigma_{\mathrm{8}}$, and $S_{\mathrm{8}}$ than the stretches. 
We note again that in these tests, we vary the mean and width independently to isolate their individual impact. In a realistic scenario, however, the uncertainties in both parameters are expected to be of similar relative magnitude. 

Additionally, we conducted tests accounting for IA and systematic effects of the source galaxy sample as de\-scribed in Sect. \ref{subs:systematics}. The inferred biases are reduced slightly by this inclusion but still significant exceeding 0.5$\,\sigma$. The results are presented in Appendix \ref{app_IA_sourcesys}.
\begin{table}
    \centering
    \caption{Biases on the 1D marginalised constraints for tests of the 3$\times$2pt analysis assuming a $\Lambda$CDM model.}
    \vspace{5mm}
   \begin{tabularx}{0.4\textwidth}{c c c c}
   \hline
    Chain & $\Omega_{\mathrm{m}}$[$\sigma$] & $\sigma_{\mathrm{8}}$[$\sigma$] & $S_{\mathrm{8}}$[$\sigma$]\\\hline
    shift 0.002(1+$z$) & 1.17 & 1.00 & 0.22 \\
    shift 0.004(1+$z$) & 2.34 & 2.05 & 0.47 \\
    stretch 1.002(1+$z$) & 0.03 & 0.02 & 0.007 \\
    stretch 1.004(1+$z$) & 0.02 & 0.002 & 0.05 \\
    stretch 1.02(1+$z$) & 0.19 & 0.24 & 0.23 \\
    stretch 1.04(1+$z$) & 0.09 & 0.24 & 0.51 \\
    \hline
\end{tabularx}
\tablefoot{In all 3$\times$2pt cases, the stretch of the distribution seems to have less of an impact compared to the photometric clustering only analysis. In contrast to that, the effect of shifts to the mean are more significant here.}
    \label{t:3x2pt sigma errors}
\end{table}

\paragraph{Photometric galaxy clustering:}
Finally, we consider the scenario of a photometric galaxy clustering analysis. We note that we also tested 2$\times$2pt setups. These are presented in Appendix \ref{app:2x2pt}.
In Fig.~\ref{gcph constraints} and Table \ref{t:GCph sigma errors} we show the constraints on $\Omega_{\rm m}$, $\sigma_{\rm 8}$, and $S_{\rm 8}$ for photometric galaxy clustering only. 
We fix the galaxy bias for diagnostic purposes to prevent it from absorbing the effects of the modifications of the redshift distributions. 
Regarding the shift values to the mean of the $n(z)$, the requirement on the sources, $\Delta z$ < 0.002\,(1+$z$) \citep{Laureijs11, EuclidSkyOverview},  holds for the lenses as well, with tolerable biases lower than 0.5\,$\sigma$ for all cosmological parameters considered.
However, we also find that the width of the distribution is important. 
In fact, as indicated by the values of the biases in the constraints listed in Table \ref{t:GCph sigma errors}, a 0.2$\%$ stretch of the distribution has a similar impact on the resulting cosmology as a shift of 0.2$\%$ corresponding to the source requirement, resulting in a bias of 0.06\,$\sigma$ in $S_{\rm 8}$. 
Nevertheless, in contrast to the analyses focussing on the 3$\times$2pt analysis, the shift of the mean seems less relevant and the effect of stretches of the distribution becomes more important. 
This effect arises because the the galaxy clustering correlation observable is essentially a local measurement, making its auto-correlation function highly sensitive to the exact radial overlap of redshift bins, and thus to the specific shape and width of the $n(z)$. 
In the full 3$\times$2pt analysis, this sensitivity is heavily diluted by the cosmic shear and galaxy-galaxy lensing signals, whose broad integration kernels along the line of sight smooth out the detailed features of the lens redshift distributions. 
\begin{table}
    \centering
    \caption{Biases on the 1D marginalised constraints obtained for photometric galaxy clustering only assuming a $\Lambda$CDM model.}
    \vspace{5mm}
   \begin{tabularx}{0.45\textwidth}{c c c c}
   \hline
    Chain & $\Omega_{\mathrm{m}}$ [$\sigma$] & $\sigma_{\mathrm{8}}$ [$\sigma$] & $S_{\mathrm{8}}$ [$\sigma$]\\\hline
    shift 0.002(1+$z$) & 0.12 & 0.15 & 0.08 \\
    shift 0.02(1+$z$) & 0.79 & 1.05 & 0.60 \\
    stretch 1 + 0.002(1+$z$) & 0.03 & 0.07 & 0.06 \\
    stretch 1 + 0.02(1+$z$) & 0.09 & 0.44 & 0.24 \\
    \hline
\end{tabularx}
\tablefoot{The galaxy bias parameters are fixed to the values provided in Table \ref{t:lambdacdm parameters} and not marginalised over in this scenario.}
    \label{t:GCph sigma errors}
\end{table}
%
%

\section{\label{sc:Conclusions}Conclusions}
In this work we studied the impact of uncertainties of the lens redshift distribution on the expected cosmological constraints obtained from \Euclid's DR1 
3$\times$2pt, 2$\times$2pt, and photometric galaxy clustering analyses.
We have investigated the loss of constraining power when marginalising over lens redshift uncertainties as well as potential biases in the constraints that arise if we assume the wrong lens redshift distribution. 
For the former, we used the FoM as a metric for the performance.
We varied both the means and widths of the $n(z)$ of the individual tomographic bins to investigate any impact of those parameters on the inferred cosmology.

First, if we analyse with the true redshift distribution but marginalise over the uncertainties on the mean and width of the $n(z)$, this leads to a loss of cons\-training power.
If we apply the default source requirement of $\sigma_{\langle z \rangle}$\,<\,0.002\,(1+$z$) to the lens redshift distribution, paired with a prior on the width that is an order of magnitude larger, the FoM degrades by 11$\%$. For $\sigma(\Delta z)$\,=\,$\sigma(\sigma_z)$\,=\,0.004\,(1+$z$), it degrades by 20$\%$ which is in accordance with the findings in \citet{kitching2008systematic}.

Furthermore, studying the impact of shifts of the mean and stretches in the width of the lens $n(z)$, we found that both perturbations bias the constraints from photometric galaxy clustering, especially the width of the distribution. 
This is in good agreement with the results from \citet{Porredon2022}, where similar cases were studied in the context of the DES-Y3 2$\times$2pt analysis.

The impact of the width of the $n(z)$ becomes less relevant when considering 3$\times$2pt scena\-rios in the $w_{\mathrm{0}}w_{a}$CDM framework and shifts of the mean become more re\-levant. 
This is also the case for scenarios in a $\Lambda$CDM framework. 

The different impact of the width uncertainties on the cosmological constraints can be explained by considering the distinct dependencies of the probes on the redshift distribution. For photometric galaxy clustering, the angular correlation function $w(\theta)$ amplitude scales inversely with the width of the redshift distribution. Approximating the distribution as a Gaussian, 
the signal scales roughly as $w(\theta) \propto \int n(z)^2 \mathrm{d}z \propto 1/\sigma_z$. 
Consequently, an underestimation of the width leads to an overestimation of the clustering amplitude, 
directly biasing the parameter $\sigma_8$. We note that while the observable statistics scale linearly with $\sigma_8$ or $S_8$ in the linear regime, the bulk of the constraining power for the other cosmological parameters originates from the shape dependence of the power spectrum as it extends into the mildly non-linear regime. In contrast to this direct dependency, cosmic shear and the galaxy-galaxy lensing signal are integrated effects dependent on the broad lensing efficiency kernel. These probes are therefore less sensitive to local perturbations in the shape of the lens redshift distribution, such as a stretch $\sigma_z$, provided the mean redshift is correctly ca\-librated. This explains why the width requirement is stringent for clustering-only analyses but becomes sub-dominant in the full 3$\times$2pt analysis, where the constraining power is driven by the combination of all three probes breaking the parameter degeneracies. While cosmic shear constrains the amplitude of the matter power spectrum, adding galaxy-galaxy lensing introduces a dependence on the galaxy bias. The degeneracy between the cosmological parameters and the galaxy bias is ultimately broken by the inclusion of galaxy clustering.  
This could cause a potential inconsistency between the probes. If not sufficiently modelled, this mismatch could manifest as a tension between the clustering and lensing parts of the data vector, introducing a bias in the joint analysis.
However, this distinct response of the probes also offers a diagnostic advantage. Since an incorrect width degrades the joint goodness-of-fit, standard internal consistency checks could serve as a validation tool to identify significant width errors before they affect the final cosmological inference.

Although we tested several scenarios, this work does not cover all aspects of the \Euclid DR1 analysis.
For instance, one caveat of our setup is the choice of the redshift distributions, which might diverge from the samples that will finally be selected for the DR1 analysis.
Moreover, we did not include the full wealth of instrumental systematic effects and focussed on selected astrophysical and observational effects.
Finally, it is challenging to fulfill the requirements for the sources already in the DR1 analysis due to the currently limited availability of spectroscopic calibration samples. 
For this reason, our analysis and results are, to some extent, not exactly representing the upcoming DR1 analysis.
We refer the reader to \citet{Blot25} and \citet{canas2025euclid} for a more extensive study of the impact of systematic effects in the context of DR3. 
While the formal requirements for DR3 will be strictly tighter due to the increased data volume, it is practically more realistic that they will be met, as we antici\-pate vastly improved spectroscopic datasets to be available by that time. 
Moreover, an initial outlook on the impact of redshift distribution uncertainties via Fisher forecast is presented in Appendix \ref{dr3fish}. Interestingly, while one might intuitively expect larger data volumes to demand strictly tighter priors, our forecast shows that the statistical power of DR3 allows the data to largely self-calibrate the nuisance parameters. Assuming perfectly known source redshifts and DR3-specific scale cuts as an idealised test case, this forecast suggests that the lens sample could tolerate uncertainties up to $\sigma (\Delta z) = \sigma (\sigma_z)$ = 0.03(1+$z$) before significantly degrading the FoM. 
We note that in a full DR3 analysis, the requirements for the source sample will be significantly more stringent. Because the lens and source redshift distributions are constrained jointly in the full analysis, the lens mean and width will easily meet their accuracy thresholds if the source galaxies meet their tighter tolerances. 
Ne\-vertheless, it will be of critical importance to ensure that the lens $n(z)$ is known with an accuracy of 0.004 (1+$z$) in the mean and width for a robust DR1 ana\-ly\-sis that can compete with stage-III results and does not critically impact \Euclid's goal of constraining $w_0$ and $w_{\rm a}$ at the 10$\%$ level. 
Thus, our work provides essential information for DR1 analysis choices. 
Given our results, we conclude that 
while the mean and width of the lens redshift distributions are intrinsically connected, the uncertainty in the mean acts as the primary limiting factor for the \Euclid DR1 cosmological analyses. 
Based on the physically motivated scenario where the uncertainties in the mean and width are of comparable order of magnitude, our tests suggest that the mean of the lens $n(z)$ must be determined to an accuracy of 0.004(1+$z$). Under these conditions, the uncertainty in the width will produce negligible additional degradation. 
This level of accuracy is required to avoid biases in the inferred cosmology larger than 0.5\,$\sigma$ and to recover 80$\%$ of the constraining power in \textit{Euclid}'s DR1 $w_0w_{a}$CDM 3$\times$2pt analysis. 
Crucially, we find that explicitly targeting this accuracy for the mean is sufficient as the impact of the width uncertainty becomes practically negligible once the mean is accurately known. 
Regarding the practical implication of these results, the derived quality selection of $\sigma(\Delta z)$\,=\,0.004\,(1+$z$) provides a clear benchmark for the DR1 lens sample. 
Since this threshold is less stringent than the 
formal weak lensing requirements targeted by the full DR3 analysis, which demand a much tighter calibration for the source sample, our findings suggest that meeting the weak lensing requirement will naturally satisfy the photometric clustering measurements as well. Therefore, for this initial data release, the galaxy clustering redshift calibration is unlikely to be the primary bottleneck for cosmological constraints, as the required precision is effectively guaranteed by the more rigorous lensing standards.

%

\begin{acknowledgements}
\AckEC \AckCosmoHub \,

KB is supported by the DLR project 50QE2305. AP acknowledges support from the European Union's Marie Skłodowska-Curie grant agreement 101068581, and from the \textit{César Nombela} Research Talent Attraction grant from the Community of Madrid (Ref. 2023-T1/TEC-29011). JF acknowledges support of Funda\c{c}\~{a}o para a Ci\^{e}ncia e a Tecnologia through the Investigador FCT Contract No. 2020.02633.CEECIND/CP1631/CT0002, the FCT project PTDC/FIS-AST/0054/2021, and the research grants UIDB/04434/2020 and UIDP/04434/2020. HHi is supported by an ERC Consolidator Grant (No. 770935) and the DLR project 50QE2305. IT has been supported by the Ramon y Cajal fellowship (RYC2023-045531-I) funded by the State Research Agency of the Spanish Ministerio de Ciencia, Innovaci\'on y Universidades, MICIU/AEI/10.13039/501100011033/, and Social European Funds plus (FSE+). IT also acknowledges support from the same ministry, via projects PID2019-11317GB, PID2022-141079NB, PID2022-138896NB; the European Research Executive Agency HORIZON-MSCA-2021-SE-01 Research and Innovation programme under the Marie Sk\l odowska-Curie grant agreement number 101086388 (LACEGAL) and the programme Unidad de Excelencia Mar\'{\i}a de Maeztu, project CEX2020-001058-M. SC acknowledges support from the Italian Ministry of University and Research (\textsc{mur}), PRIN 2022 `EXSKALIBUR – Euclid-Cross-SKA: Likelihood Inference Building for Universe's Research', Grant No.\ 20222BBYB9, CUP D53D2300252 0006, from the Italian Ministry of Foreign Affairs and International
Cooperation (\textsc{maeci}), Grant No.\ ZA23GR03, and from the European Union -- Next Generation EU.\\
The analysis in this work was done with \texttt{Python} (\url{http://www.python.org}), including the packages \texttt{NumPy} \citep{numpy} and \texttt{SciPy} \citep{2020SciPy-NMeth}. The plots
have been generated using \texttt{Matplotlib} \citep{matplotlibHunter:2007} and \texttt{Chainconsumer} \citep{Chainconsumer}.
\end{acknowledgements}

%
%

\bibliography{my, Euclid}

%

 \begin{appendix}

\section{Application to Flagship measurements}\label{app_flagship}

To go beyond the previously used synthetic data vectors, we also consider the scenario of photometric galaxy clustering with a data vector, the 2-point angular correlation function, measured on the Flagship simulation with the Landy--Szalay estimator \citep{Landy_1993}
\begin{equation}
w(\theta)=\frac{\NDD-2\NDR+\NRR}{\NRR}\, ,
\end{equation}
where \NDD, \NDR, and \NRR\, are the pair counts with D represen\-ting the data and R a random point. The random catalogues are created by sampling the footprint of the simulation defined by a \texttt{HEALPix} mask of $N_\mathrm{side} = 4096$ (angular resolution $\Delta\theta=
\ang{0.014;;}$). The ratio between the number of random points and galaxies is 30, which yields a total of $103\expo{6}$ points per redshift bin. We use 20 logarithmic angular bins between $\theta_\mathrm{min}=\ang{;2.5;}$ and $\theta_\mathrm{max}=\ang{;250}$. The measurement is performed with the code \texttt{TreeCorr} \citep{Jarvis_2004}.

The theoretical model 
is computed following Eq. \eqref{eq:w_theta} with the fiducial cosmology of the simulation, detailed in Table \ref{t:lambdacdm parameters}. Moreover, this model assumes a linear galaxy bias which is fitted to the measured data vector. This fit is performed with the theoretical model computed assuming a galaxy bias $b=1$ and the scales used are  $6\arcmin\leq\theta\leq60\arcmin$. The linear galaxy bias obtained from the fit in the six redshift bins is \{1.144, 1.259, 1.292, 1.407, 1.455, 1.720\}. The small differences with respect to the values listed in Table~\ref{t:lambdacdm parameters} are explained by the fact that the values of the table were inferred from a smaller area of 400 deg$^2$. Moreover, we use $m_\nu=0.06\;\mathrm{eV}$, unlike the analyses with the theory data vectors which used $m_\nu=0\;\mathrm{eV}$. For this reason, the associated fiducial values of $\sigma_8$ and $S_8$ shift from 0.828 and 0.854 to 0.817 and 0.843, respectively.

The values in Table~\ref{t:lambdacdm parameters} were derived from 400 deg$^2$ to match the choices made in the internal \Euclid science performance verification 3 exercise. However, in this complementary analysis, the fiducial values were adapted in the model to best match the properties of the Flagship simulation area on which the data vectors were measured, for two main reasons. 
First, using the $n(z)$ or galaxy bias measured in the small area would significantly bias the model due to sample variance when we compare it to the measurements from the 2500\,$\deg^2$ area. Second, measuring $w(\theta)$ in a small area yields a measurement with much higher statistical noise, as expected from the difference in the galaxy number density.

We find significant deviations between the model and measurement in the first redshift bin. Because the single realisation of the Flagship simulation available inherently exhibits a lack of clustering at $\theta>\ang{;50}$ in this specific bin, we provide the baseline constraints both with and without it to ensure robustness.

\begin{table*}
\centering
\caption{Constraints from photometric galaxy clustering alone with the measured data vector  as a function of the scale cuts used in the baseline analysis.}
\begin{tabular}{cccccccc}
\hline
Scale cuts & $\Omega_\mathrm{m}\;[\sigma]$   & $h\;[\sigma]$   & $S_8\;[\sigma]$   & $\Omega_\mathrm{b}\;[\sigma]$   & $n_\mathrm{s}\;[\sigma]$   & $A_\mathrm{s}\;[\sigma]$   & $\sigma_8\;[\sigma]$   \\
\hline
 4 Mpc/$h$  & $0.28$  & $0.45$& $0.14$  & $0.45$  & $1.41$  & $0.62$  & $2.18$\\
 4 Mpc/$h$, no bin 1   & $0.77$  & $0.83$& $0.33$  & $0.56$  & $1.75$  & $0.89$  & $2.65$\\
 8 Mpc/$h$   & $0.66$  & $0.38$& $0.92$  & $1.82$  & $0.33$  & $0.83$  & $0.87$\\
 8 Mpc/$h$, no bin 1   & $1.16$  & $0.64$& $1.47$  & $2.10$  & $0.12$  & $0.89$  & $0.89$\\
 8 Mpc/$h$, 100 arcmin      & $0.25$  & $0.26$& $0.09$  & $0.14$  & $0.03$  & $0.07$  & $0.66$\\
 8 Mpc/$h$, 100 arcmin, no bin 1 & $0.08$  & $0.08$& $0.30$  & $0.14$  & $0.10$  & $0.14$  & $0.77$\\
\hline
\end{tabular}
\label{t:shifts_baseline_scale_cuts_flagship}
\end{table*}

We use the Monte Carlo method to sample the parameter space with the measured data vector. The same covariance and methodology are used as in Sect.~\ref{sc:Method}. 
Given that we use a linear model for the galaxy bias, we adopt conservative scale cuts of $R$ = 8\,Mpc\,$h^{-1}$ in Eq. \eqref{eq:scale_cut} and $\theta_\mathrm{max}=\ang{;100}$, following the standards set by the DES-Y3 photometric galaxy clustering analysis \citep{krause2017dark,Porredon2022}. 
All shifts in the baseline analysis depending on the scale cuts used are grouped in Table \ref{t:shifts_baseline_scale_cuts_flagship}. 
It should be noted that the theoretical model for $w(\theta)$ exhibits a baseline mismatch with the specific clustering properties of this Flagship realisation. Consequently, this fixed-bias test serves purely as an isolated diagnostic for $n(z)$ variations, and the absolute bias values, which increase under scale cuts due to this underlying model-mock discrepancy, do not directly translate to the fully marginalised cosmological constraints. 
We also include constraints excluding the first redshift bin, given the significant differences between measurement and model. 
While our primary focus is on the differential impact of $n(z)$ errors that can be evaluated even in the presence of model biases, we adopt these conservative cuts to ensure a stable baseline where the linear bias model is valid and to compare against the results of the analysis with simulated theory data vectors. This guarantees that the parameter shifts discussed below are driven by $n(z)$ uncertainties and not strongly coupled with non-linear modelling issues.
With $R$ = 8\,Mpc\,$h^{-1}$, the biases on $\sigma_8$ and $n_\mathrm{s}$ decrease to $0.87\,\sigma$ and $0.33\,\sigma$, but the bias on $\Omega_\mathrm{b}$ reaches $1.82\,\sigma$. Only with both scale cuts ($R$ = 8\,Mpc\,$h^{-1}$,\, $\theta_\mathrm{max}=\ang{;100}$) can we limit the biases to be smaller than $1\,\sigma$ without applying informative priors. In this last case, removing the first redshift bin decreases the bias in $\Omega_\mathrm{m}$ and $h$ but also increases the other biases. With less strict scale cuts, removing this bin generally increases biases.

\begin{figure}
     \centering
     \includegraphics[width=0.45\textwidth]{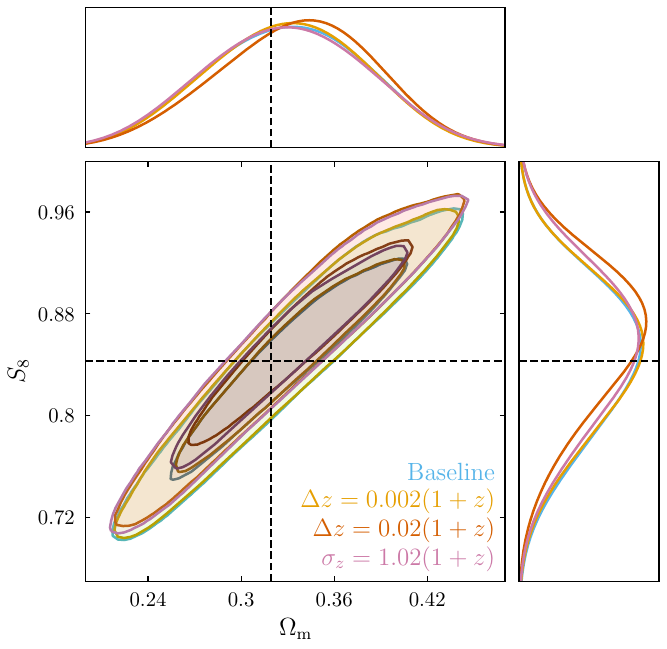}
     \caption{Comparison of cosmological constraints on $\Omega_{\mathrm{m}}$ and $S_{\mathrm{8}}$ considering photometric galaxy clustering only in flat $\Lambda$CDM obtained from measured data vectors for different shifts and stretches of the lens redshift distribution. The two sets of contours indicate the 68$\%$ and 95$\%$ confidence levels.}
     \label{fig:gcph constraints flagship}
\end{figure}

\begin{table}
\centering
\caption{Biases of the contours obtained for the tests on photometric
galaxy clustering only measured on the Flagship simulation and with fixed galaxy bias.}
\vspace{5mm}
   \begin{tabularx}{0.38\textwidth}{c c c c}
\hline
Chain & $\Omega_\mathrm{m}\;[\sigma]$  & $\sigma_8\;[\sigma]$ &  $S_8\;[\sigma]$ \\\hline
shift 0.002(1+$z$) & 0.04 & 0.03 & 0.04 \\
shift 0.02(1+$z$) & 0.24 & 0.39 & 0.37 \\
stretch 1.02(1+$z$) & 0.02 & 0.44 & 0.13  \\
\hline
\end{tabularx}
\label{t:shifts_lcdm_flagship}
\end{table}

In Fig.~\ref{fig:gcph constraints flagship}, we show the constraints on $S_8$ and $\Omega_\mathrm{m}$ in the various cases of a biased redshift distribution with $\Delta z=0.002(1+z)$, a strongly biased distribution with $\Delta z=0.02(1+z)$, and a stretched distribution with $\sigma_z=1.02(1+z)$. The associated shifts on $\Omega_\mathrm{m}$, $\sigma_8$, and $S_8$ are listed in Table \ref{t:shifts_lcdm_flagship}. We find that the smallest bias of the redshift distribution has very little impact over the cosmological constraints, with a maximum shift of $0.04\,\sigma$ on $S_8$ and $\Omega_\mathrm{m}$ with respect to the baseline. This impact is consistent with the one obtained with the theoretical data vectors (see Table \ref{t:GCph sigma errors}). However, the larger bias $\Delta z=0.02(1+z)$ induces a shift of $0.39\,\sigma$ on $\sigma_8$, $0.24\,\sigma$ on $\Omega_\mathrm{m}$, and $0.37\,\sigma$ on $S_8$, respectively. In this case, the bias on $\Omm$ is $0.55\,\sigma$ smaller than with theoretical data vectors and $0.66\,\sigma$ smaller for $\sigma_8$. As for the stretch $\sigma_z=1.02(1+z)$, it causes a slightly larger shift on $\sigma_8$ with $0.44\,\sigma$, a smaller shift of $0.13\,\sigma$ on $S_8$, and $0.02\,\sigma$ on $\Omega_\mathrm{m}$, which is similar to the results from Sect.~\ref{biasess}.

 \section{Comparison between the Monte Carlo results and a Fisher matrix approximation}\label{app:comp_fish}

Here we compare the Fisher matrix analyses with the results from the Monte Carlo simulations. We run the Fisher forecast as described in Sect.~\ref{subsec:likefish} and estimate the bias in the best-fit value. 

\begin{figure}[h!]
\centering
\includegraphics[angle=0,width=1.0\hsize]{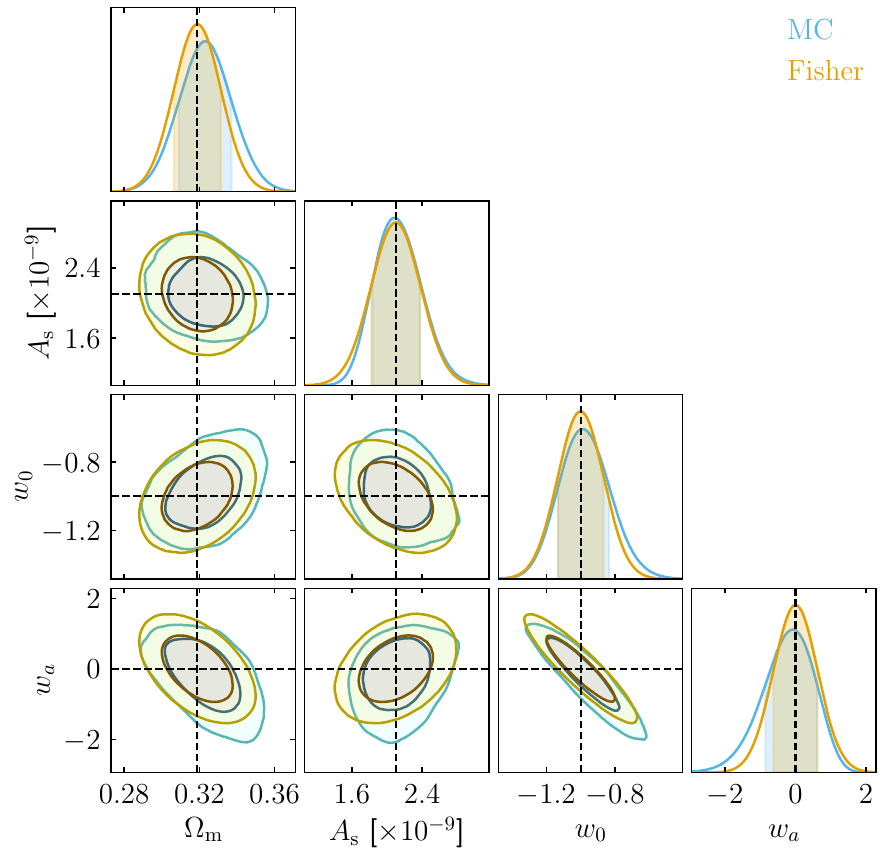}
\caption{Comparison of cosmological constraints for $\Omega_{\mathrm{m}}$, $A_{\rm s}$, $w_0$, and $w_a$ between Monte Carlo chains and Fisher forecasts for the baseline setting. The contours indicate the 68$\%$ and 95$\%$ confidence level.}
\label{fig:compare_chains_fisher}
\end{figure}

 We start by comparing the Fisher outcomes with the baseline Monte Carlo results. We present the comparison in Fig.~\ref{fig:compare_chains_fisher}. As expected, the Fisher matrix provides more stringent constraints, and perfect ellipses by construction, but overall they seem to be in agreement with the Monte Carlo posteriors. 

\begin{figure}[h!]
\centering
\includegraphics[angle=0,width=1.0\hsize]{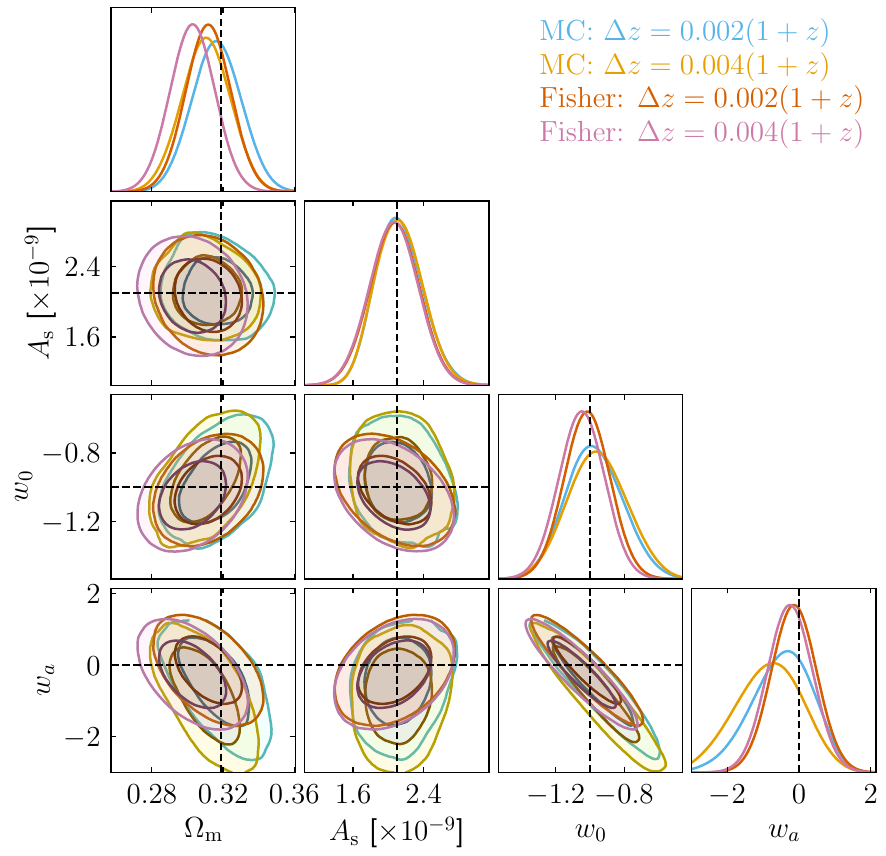}
\includegraphics[angle=0,width=1.0\hsize]{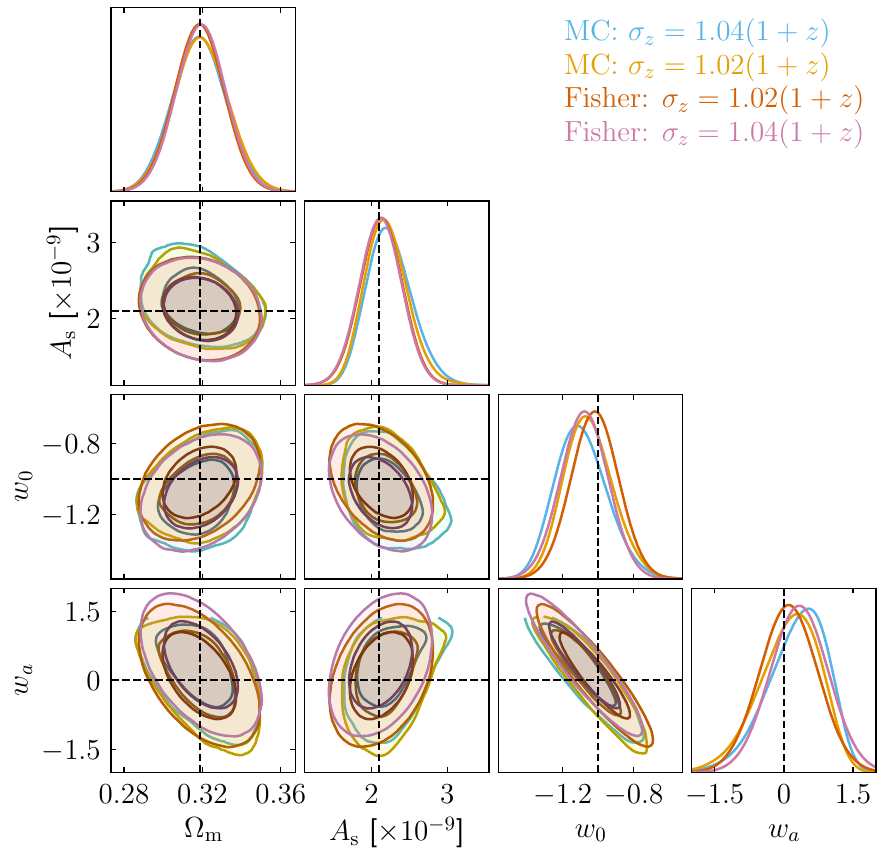}
\caption{Comparison of cosmological constraints for $\Omega_{\mathrm{m}}$, $A_{\rm s}$, $w_0$, and $w_a$ between Monte Carlo chains and Fisher forecasts. The contours indicate the 68$\%$ and 95$\%$ confidence level. {\em Top}: biases due to shifting the mean of the redshift distributions. {\em Bottom}: biases due to stretching the width of the redshift distributions.}
\label{fig:compare_chains_fisher_biasdistrib}
\end{figure}

We then compute the expected shifts in the best-fit values of the cosmological parameters resulting from assuming the wrong redshift distributions in the individual tomographic bins. In the top panel of Fig.~\ref{fig:compare_chains_fisher_biasdistrib}, we compare the Monte Carlo and the Fisher predictions in two cases of shifted bin centres. The estimates with the Fisher matrix, while not as accurate, display a similar trend and order of magnitude of the shifts in the best-fit value compared to those estimated in the Monte Carlo analysis. A similar conclusion can be drawn if the redshift distributions are stretched (bottom panel of Fig.~\ref{fig:compare_chains_fisher_biasdistrib}).

\begin{figure}
\centering
\includegraphics[angle=0,width=1.0\hsize]{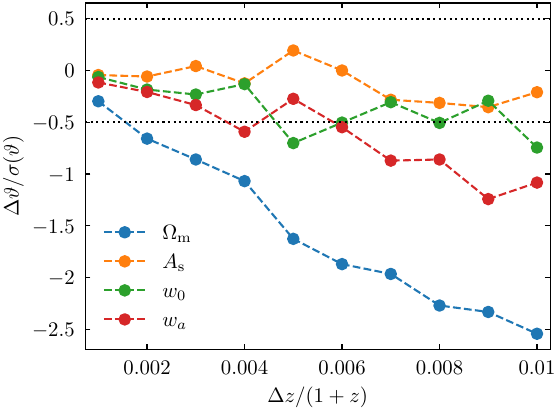}
\includegraphics[angle=0,width=1.0\hsize]{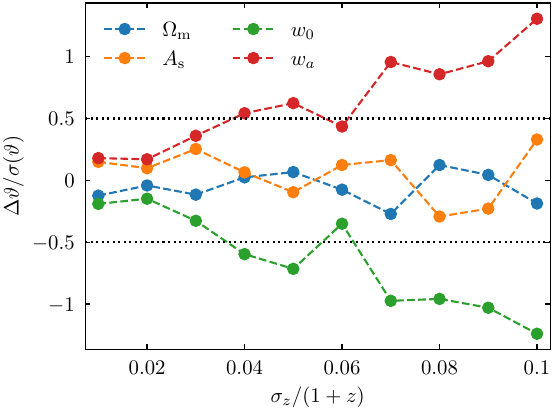}
\caption{Relative best-fit bias for $\Omega_{\mathrm{m}}$, $A_{\rm s}$, $w_0$, and $w_a$ as a function of {\em top:} shifts in the central bin redshift; {\em bottom:} widening of the bin distribution width.}
\label{fig:extrapolation_shifts}
\end{figure}

\newpage
Moreover, we extrapolate the effect of more extreme shifts and stretches on the best fit of the cosmological parameters that we present in Fig.~\ref{fig:extrapolation_shifts}. One should note that such estimates are based on a Gaussian approximation of the posterior and linear perturbations around the maximum of the posterior. Therefore, once $\Delta \theta/\theta \sim 1$, we expected our approximation to break down. Still, it gives a qualitative estimate of the expected shifts. The results show that uncertainties in the mean of the lens redshift distributions will bias the cosmological parameters in line with Sect.~\ref{biasess}. 

\newpage

\section{Impact of IA and systematic effects of the source galaxy sample}\label{app_IA_sourcesys}
Additionally, we conducted tests accounting for IA and systematic effects of the source galaxy sample as de\-scribed in Sect. \ref{subs:systematics}. 
For these cases, we generated a new data vector with the fiducial values for IA and marginalised over both IA and weak lensing nuisance parameters in the likelihood analysis (values and priors in Table \ref{t:lambdacdm parameters}).
The biases we obtain 
are reduced slightly by the inclusion of the systematic effects (e.g., 30$\%$ smaller in $\Omega_{\mathrm{m}}$ if IA and source systematic effects are included) but are still very large (i.e., greater than 0.5\,$\sigma$).
\newpage
\begin{table}
    \centering
    \caption{Biases on the 1D marginalised constraints for selected tests of the 3$\times$2pt analysis. 
    }
    \vspace{5mm}
   \begin{tabularx}{0.45\textwidth}{c c c c}
   \hline
    Chain & $\Omega_{\mathrm{m}}$[$\sigma$] & $\sigma_{\mathrm{8}}$[$\sigma$] & $S_{\mathrm{8}}$[$\sigma$]\\\hline
    & \textbf{No IA} & \\
    shift 0.002(1+$z$) & 1.17 & 1.00 & 0.22 \\
    shift 0.004(1+$z$) & 2.34 & 2.05 & 0.47 \\
    stretch 1.02(1+$z$) & 0.19 & 0.24 & 0.23 \\
    stretch 1.04(1+$z$) & 0.09 & 0.24 & 0.51 \\
    \hline
    & \textbf{With IA} & \\
    shift 0.002(1+$z$) & 0.77 & 0.68 & 0.20 \\
    shift 0.004(1+$z$) & 1.51 & 1.37 & 0.37 \\
    stretch 1.02(1+$z$) & 0.13 & 0.06 & 0.23 \\
    stretch 1.04(1+$z$) & 0.52 & 0.35 & 0.49 \\
    \hline
    & \textbf{IA and source sys} & \\
    shift 0.002(1+$z$) & 0.80 & 0.70 & 0.22 \\
    shift 0.004(1+$z$) & 1.52 & 1.37 & 0.45 \\
    stretch 1.02(1+$z$) & 0.12 & 0.04 & 0.26 \\
    stretch 1.04(1+$z$) & 0.55 & 0.37 & 0.50 \\
    \hline
\end{tabularx}
\tablefoot{We explored several scenarios including IA and source systematic effects (photo-$z$ uncertainties and multiplicative shear bias). In all 3$\times$2pt cases, the stretch of the distribution seems to have less of an impact compared to the photometric clustering only analysis. In contrast to that, the effect of shifts to the mean are more significant here.}
    \label{t:3x2pt sigma errors_IA}
\end{table}
\newpage

\section{\label{app:2x2pt} Results from 2$\times$2pt tests}
We conducted several tests in a 2$\times$2pt scenario (i.e., the combination of photometric galaxy clustering and galaxy-galaxy lensing).
Given that coherent modifications of the redshift distributions are likely not the most realistic scenario, we conducted several tests for alternating shifts/stretches in a 2$\times$2pt scenario. 
That is, in this case the uneven bins were shifted/stretched in the positive direction, whereas negative shifts/stretches were applied to the even bins. 
Besides exploring the impact of alternating shifts/stretches, we tested a scenario in which both lens $n(z)$ parameters, shift and stretch, were modified simultaneously (in the same direction for all bins).
This is likely the most realistic case, as both parameters will come with some uncertainty in the upcoming real-data analysis.
In contrast to the clustering only analysis, the galaxy bias was free and marginalised over.
The resulting biases are listed in Table \ref{t:2x2pt sigma errors}.

\begin{table}[h!]
    \centering
    \caption{Biases of the constraints obtained for the tests of the 2$\times$2pt scenario. 
    }
    \vspace{5mm}
   \begin{tabularx}{0.45\textwidth}{c c c c}
   \hline
    Case & $\Omega_{\mathrm{m}}$[$\sigma$]& $\sigma_{\mathrm{8}}$[$\sigma$]& $S_{\mathrm{8}}$[$\sigma$]\\\hline
    & \textbf{Coherent} & \\
    shift 0.002(1+$z$) & 0.92 & 0.75 & 0.12 \\
    shift 0.004(1+$z$) & 1.84 & 1.50 & 0.16 \\
    stretch 1.02(1+$z$) & 0.18 & 0.40 & 0.62 \\
    stretch 1.04(1+$z$) & 0.48 & 0.90 & 1.24 \\
    \hline
    & \textbf{Alternating} & \\
    shift +/- 0.002(1+$z$) & 0.024 & 0.14 & 0.35 \\
    shift +/- 0.004(1+$z$) & 0.071 & 0.26 & 0.68 \\
    stretch 1.0 +/- 0.02(1+$z$) & 0.55 & 0.52 & 0.26 \\
    stretch 1.0 +/- 0.04(1+$z$) & 1.30 & 1.26 & 0.58 \\
    \hline 
    & \textbf{Both} & \\
    shift 0.002(1+$z$),  & \multirow{2}{0.75cm}{1.45} & \multirow{2}{0.75cm}{1.29} &\multirow{2}{0.75cm}{0.39}\\
    stretch 1.02(1+$z$) & & \\
    \hline
\end{tabularx}
\tablefoot{The tests include coherent modifications as well as alternating shifts/stretches and a case in which both $n(z)$ parameters are modified.}
    \label{t:2x2pt sigma errors}
\end{table}

\begin{figure}
    \centering
    \includegraphics[width=1.0\hsize]{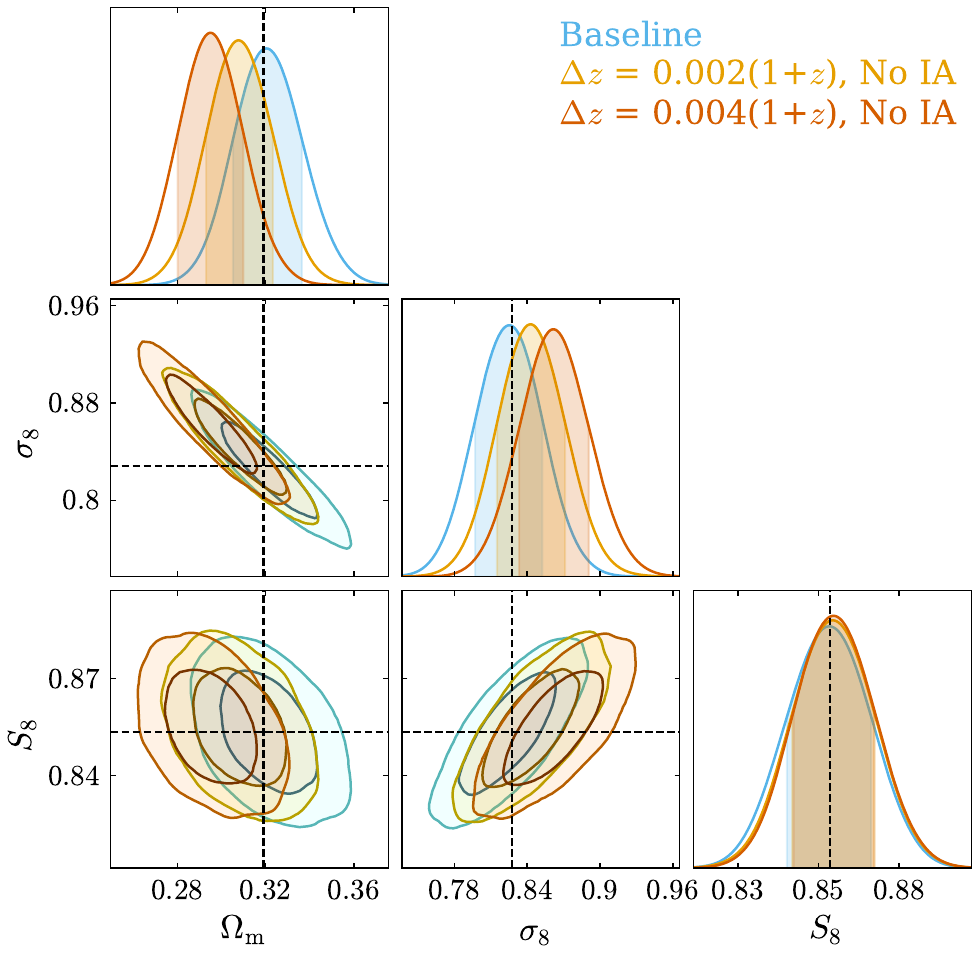}
    \caption{Comparison of cosmological constraints on $\Omega_{\mathrm{m}}$, $\sigma_{\mathrm{8}}$, and $S_{\mathrm{8}}$ in $\Lambda$CDM obtained from the baseline and the ones obtained from a coherently-shifted distribution in a 2$\times$2pt scenario.}
    \label{fig:2x2pt_bias}
\end{figure}
\begin{figure}
    \centering
    \includegraphics[width=1.0\hsize]{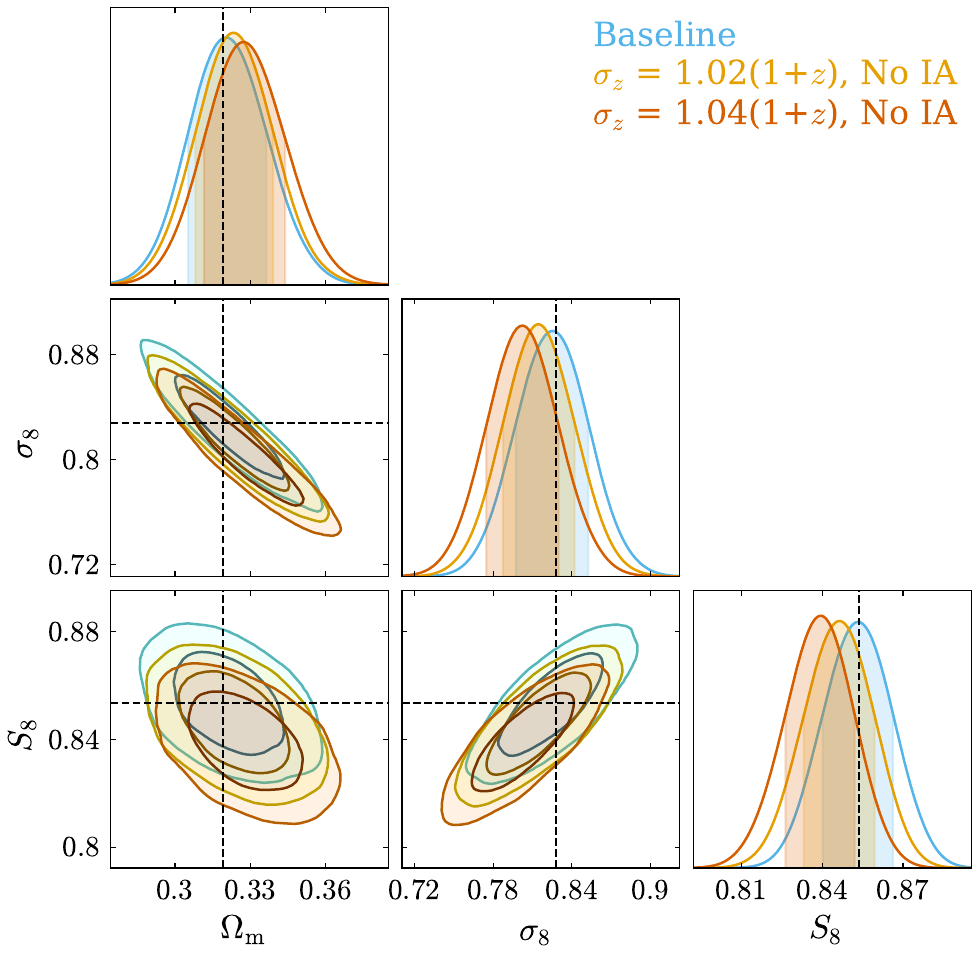}
    \caption{Comparison of cosmological constraints on $\Omega_{\mathrm{m}}$, $\sigma_{\mathrm{8}}$, and $S_{\mathrm{8}}$ in $\Lambda$CDM obtained from the baseline and the ones obtained from a coherently-stretched distribution in a 2$\times$2pt scenario.}
    \label{fig:2x2pt_width}
\end{figure}

\paragraph{Coherent modifications:}The 2$\times$2pt analysis includes the galaxy-galaxy lensing 2-point correlation function that depends less on the exact shape of the lens $n(z)$ and is instead also affected by the source sample's lensing efficiency.
Again, we neglect uncertainties in the source redshift distribution and assume it to be perfectly known.
This reduces the impact of the stretches and enhances the relevance of the mean shifts.
As indicated by the contour plots from Fig.~\ref{fig:2x2pt_width}, the impact of shifts to the mean is greater than in the galaxy clustering only analysis (with the bias in $\sigma_{\mathrm{8}}$ being about 80$\%$ smaller in the clustering analysis), while the biases on the cosmology are reduced (e.g., 10$\%$ smaller in $\sigma_{\mathrm{8}}$) if the distribution is stretched.

\begin{figure}
    \centering
    \includegraphics[width = 1.0\hsize]{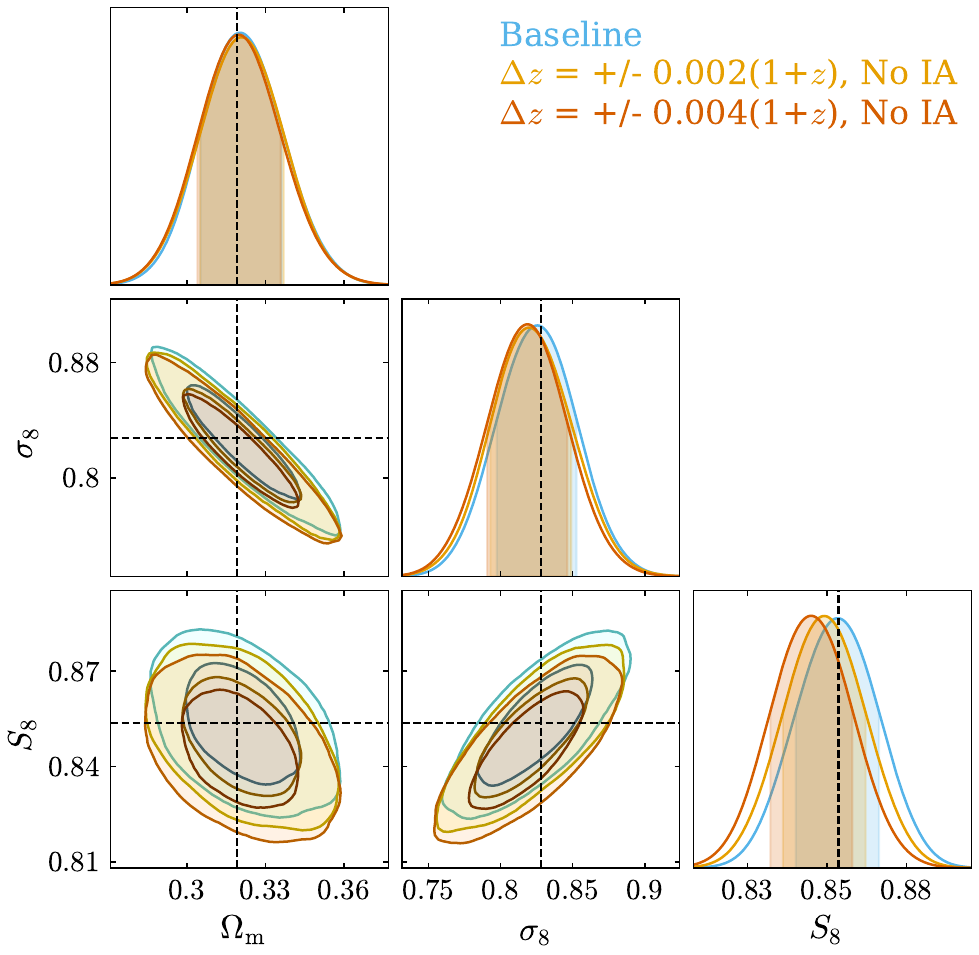}
    \caption{
    Same as Fig. \ref{fig:2x2pt_bias}, now the shifts applied alternate every bin.}
    \label{fig:2x2pt_alt_bias}
\end{figure}
\begin{figure}
    \centering
    \includegraphics[width=1.0\hsize]{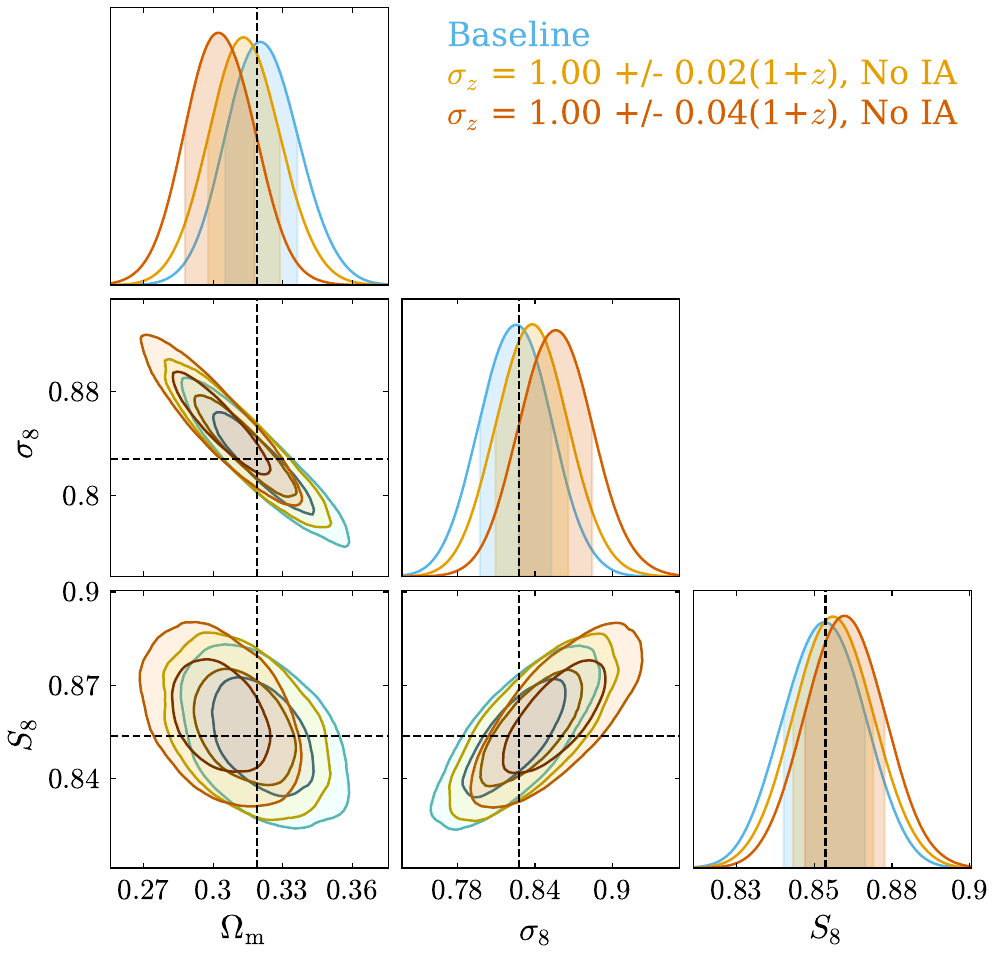}
    \caption{
    Same as Fig. \ref{fig:2x2pt_width}, now the stretches applied alternate every bin.}
    \label{fig:2x2pt_alt_width}
\end{figure}

\paragraph{Alternating modifications:}The previously studied coherent modifications of the redshift distributions are likely not the most realistic scenario.
Therefore, we conducted several tests applying alternating shifts/stretches.
The uneven bins were shifted/stretched in positive direction, while negative shifts/stretches were applied to the even bins.
The results (see Figs. \ref{fig:2x2pt_alt_bias} and \ref{fig:2x2pt_alt_width}) show that this approach reduces the impact of the mean shifts.\\

\begin{figure}
    \centering
    \includegraphics[width=1.0\hsize]{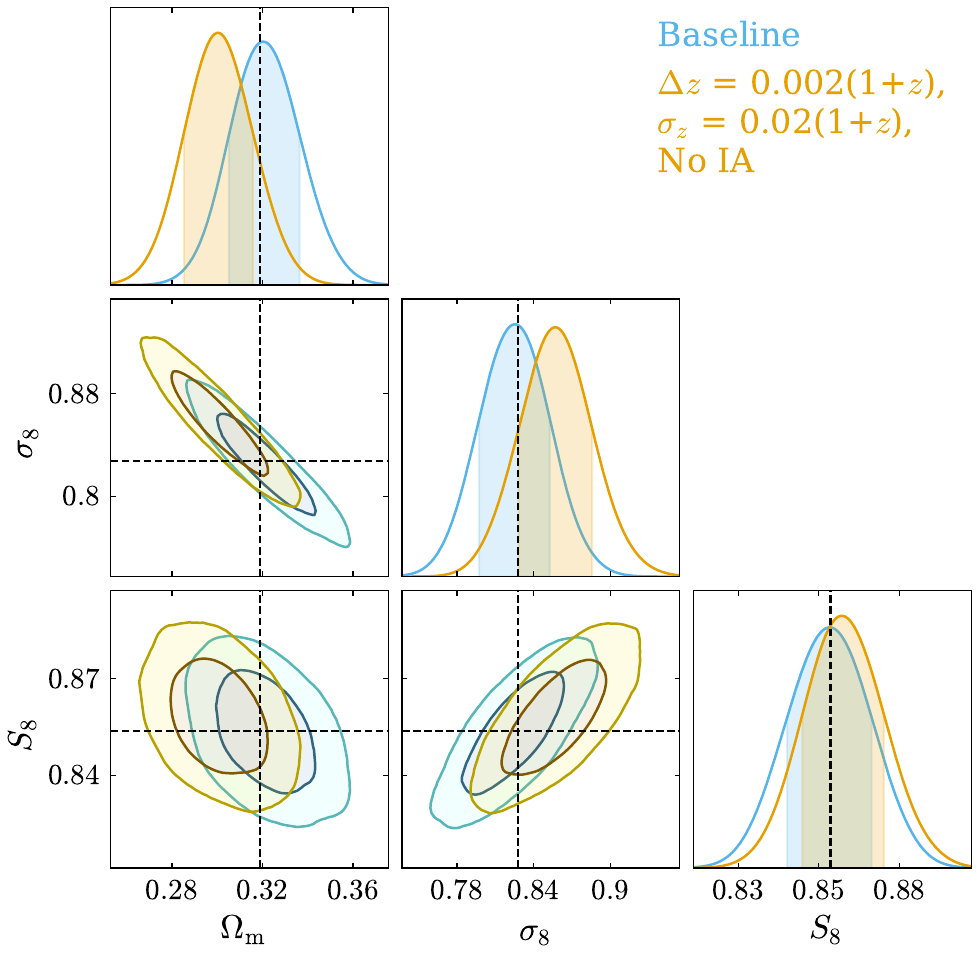}
    \caption{Comparison of cosmological constraints 
    obtained from a shifted and stretched distribution in a 2$\times$2pt scenario.}
    \label{fig:2x2pt_both}
\end{figure}

\paragraph{Modifying both parameters:}Finally, we considered one scenario in which both lens $n(z)$ parameters were modified simultaneously.
The results, presented in Fig.~\ref{fig:2x2pt_both}, clearly show that the biases in the cosmological parameters are larger than those caused by individual modification, with biases of over 1\,$\sigma$ in both $\Omega_{\mathrm{m}}$ and $\sigma_{\mathrm{8}}$. 
This emphasises the need to pin both $n(z)$ parameters down to sufficient precision.

\section{Outlook to DR3}\label{dr3fish}

As an initial outlook on the forthcoming DR3, we conduct Fisher forecasts to evaluate the sensitivity to redshift distribution uncertainties following the methodology presented in Sect. \ref{sbc:explo_fisher}. We adopt the fiducial cosmological parameters and redshift distributions from \citet{EuclidSkyOverview}, and compute a DR3-like covariance matrix using \texttt{CosmoCov}. While the specific analysis choices such as scale cuts may differ from those in \citet{EuclidSkyOverview}, these forecasts serve as a robust preliminary assessment of how redshift distribution uncertainties will impact DR3.
The FoM percentages with respect to the baseline case are listed in Table \ref{t:FoMdr3}.
In this case, 80$\%$ of the baseline FoM are recovered for $\sigma (\Delta z) = \sigma (\sigma_z)$ = 0.03(1+$z$).
A more comprehensive analysis of these sensitivities is deferred to future work.
\begin{table}
    \centering
    \caption{FoM percentages for a DR3 scenario recovered using Fisher forecast.}
    \vspace{5mm}
   \begin{tabularx}{0.45\textwidth}{c c }
   \hline
    Chain & $w_{\mathrm{0}}\times w_{a}$ 
    \\\hline
    $\sigma (\Delta z)$ = 0.001(1+$z$), $\sigma (\sigma_z)$ = 0.001(1+$z$) & 90.1\\
    $\sigma (\Delta z)$ = 0.002(1+$z$), $\sigma (\sigma_z)$ = 0.002(1+$z$) & 88.5\\
    $\sigma (\Delta z)$ = 0.004(1+$z$), $\sigma (\sigma_z)$ = 0.004(1+$z$) & 87.1\\
    $\sigma (\Delta z)$ = 0.008(1+$z$), $\sigma (\sigma_z)$ = 0.008(1+$z$) & 85.0\\
    $\sigma (\Delta z)$ = 0.01(1+$z$), $\sigma (\sigma_z)$ = 0.01(1+$z$) & 84.1\\
    $\sigma (\Delta z)$ = 0.02(1+$z$), $\sigma (\sigma_z)$ = 0.02(1+$z$) & 81.3\\
    $\sigma (\Delta z)$ = 0.03(1+$z$), $\sigma (\sigma_z)$ = 0.03(1+$z$) & 80.3\\
    \hline
    \hline
    \hline
    \end{tabularx}
    \tablefoot{The FoMs were obtained in the 3$\times$2pt analysis of a simulated theory data vector with the true redshift distribution, marginalising over uncertainties on the mean ($\Delta z$) and width ($\sigma_z$) of the lens redshift distributions and assuming different standard deviations $\sigma$ of the corresponding Gaussian priors. The numbers quoted are the percentage of the FoM recovered  with respect to the \lq baseline\rq case that corresponds to the non-marginalisation of lens photo-$z$ parameters.}
    \label{t:FoMdr3}
\end{table}

 \end{appendix}

\end{document}